\documentclass[journal]{IEEEtran}

\usepackage{algorithm2e}
\usepackage{bm}
\usepackage{footnote}
\usepackage{upgreek}
\usepackage{epsf,exscale,times}
\usepackage{graphicx}
\usepackage{amssymb}
\usepackage{amsmath}
\usepackage{amsthm}
\usepackage{amsfonts}
\usepackage{fancyhdr}
\usepackage{subcaption}
\usepackage{tikz}
\usepackage{multirow}
\usepackage{multicol}
\usepackage{glossaries}
\usepackage{cite}
\usepackage{nccmath}
\usepackage{commath}
\usepackage{comment}
\usepackage{color}
\usepackage{mathtools}
\usepackage{booktabs}
\usepackage{siunitx}
\usepackage{hyperref}

\newcommand{\bA}{\mbox{\boldmath{$A$}}}
\newcommand{\ba}{\mbox{\boldmath{$a$}}}

\newcommand{\bF}{\mbox{\boldmath{$F$}}}

\newcommand{\bh}{\mbox{\boldmath{$h$}}}

\newcommand{\bX}{\mbox{\boldmath{$X$}}}
\newcommand{\bx}{\mbox{\boldmath{$x$}}}

\newacronym{APAS}{APAS}{Almost Perfect auto-correlation Sequences}
\newacronym{MPS}{MPS}{Minimum Peak Sidelobe} 
\newacronym{PMCW}{PMCW}{Phase Modulated Continuous Wave}
\newacronym{MF}{MF}{Merit Factor}
\newacronym{SNR}{SNR}{Signal to Noise Ratio}
\newacronym{INR}{INR}{Interference to Noise Ratio}
\newacronym{SINR}{SINR}{Signal to Interference plus Noise Ratio}
\newacronym{AF}{AF}{Ambiguity Function}
\newacronym{MIMO}{MIMO}{Multiple Input Multiple Output}
\newacronym{SISO}{SISO}{Single Input Single Output}
\newacronym{CD}{CD}{Coordinate Descent}
\newacronym{BCD}{BCD}{Block Coordinate Descent}
\newacronym{GD}{GD}{Gradient Descent}
\newacronym{MM}{MM}{Majorization-Minimization}
\newacronym{FMCW}{FMCW}{Frequency Modulated Continuous Wave}
\newacronym{CDM}{CDM}{Code Division Multiplexing}
\newacronym{DFT}{DFT}{Discrete Fourier Transform}
\newacronym{FFT}{FFT}{Fast Fourier Transform}
\newacronym{MVDR}{MVDR}{Minimum Variance Distortionless Response}
\newacronym{MBI}{MBI}{Maximum Block Improvement}
\newacronym{RFPA}{RFPA}{Radio Frequency Power Amplifier}
\newacronym{BPSK}{BPSK}{Binary Phase Shift Keying}
\newacronym{QPSK}{QPSK}{Quadrature Phase Shift Keying}
\newacronym{ULA}{ULA}{Uniform Linear Array}
\newacronym{DOF}{DOF}{Degrees of Freedom}
\newacronym{PSK}{PSK}{Phase Shift Keying}
\newacronym{PSL}{PSL}{Peak Sidelobe Level}
\newacronym{ISL}{ISL}{Integrated Sidelobe Level}
\newacronym{ISLR}{ISLR}{Integrated Sidelobe Level Ratio}
\newacronym{SILR}{SILR}{Spectral Integrated Level Ratio}
\newacronym{LFM}{LFM}{Linear Frequency Modulation}
\newacronym{HPM}{HPM}{Hybrid Phased MIMO}
\newacronym{MPSK}{MPSK}{$M$-ary Phase Shift Keying}
\newacronym{LPI}{LPI}{Low Probability of Intercept}
\newacronym{RoC}{RoC}{Radar-on-Chip}
\newacronym{RF}{RF}{Radio-Frequency}
\newacronym{PAR}{PAR}{Peak-to-Average Power Ratio}
\newacronym{LTE}{LTE}{Long Term Evolution}
\newacronym{DL}{DL}{Down Link}
\newacronym{UL}{UL}{Up Link}
\newacronym{iid}{i.i.d.}{independent and identically distributed}
\newacronym{BS}{BS}{Base Station}
\newacronym{BSUM}{BSUM}{Block Successive Upper-bound Minimization}
\newacronym{SDR}{SDR}{Software-Defined Radio}
\newacronym{OTA}{OTA}{Over-The-Air}
\newacronym{USRP}{USRP}{Universal Software Radio Peripheral}
\newacronym{ICCL}{ICCL}{Integrated Cross Correlation Level}
\newacronym{ADMM}{ADMM}{Alternating Direction Method of Multipliers}
\newacronym{SDPM}{SDPM}{Spectrum Discrete Phase Modulation}
\newacronym{CW}{CW}{Continuous Wave}
\newacronym{DoA}{DoA}{Direction of Arrival}
\newacronym{MUSIC}{MUSIC}{Multiple Signal Classification}
\newacronym{BiST}{BiST}{Binary Sequences seTs}
\newacronym{PDSCH}{PDSCH}{Physical Downlink Shared Channel}
\newacronym{PDCCH}{PDCCH}{Physical Downlink Control Channel}
\newacronym{MCS}{MCS}{Modulation and Coding Schemes}
\newacronym{GUI}{GUI}{Graphical User Interface}
\newacronym{MI}{MI}{mutual information}
\newacronym{NI}{NI}{National Instruments}
\newacronym{HW}{HW}{hardware}
\begin{document}
\title{Coexistence of Communications and Cognitive MIMO Radar: Waveform Design and Prototype}
%
%
%
\author{
        Mohammad~Alaee-Kerahroodi,~\IEEEmembership{Member,~IEEE,}
Ehsan~Raei,~\IEEEmembership{Student Member,~IEEE,}
        Sumit~Kumar,~\IEEEmembership{Member,~IEEE},
        Bhavani~Shankar~Mysore~Rama~Rao,~\IEEEmembership{Senior~Member,~IEEE}
}
%
%
\maketitle
%
\begin{abstract}
New generation of radar systems will need to coexist with other radio frequency (RF) systems, anticipating their behavior and  reacting appropriately to avoid interference. In light of this requirement, this paper designs, implements, and evaluates the performance of phase-only sequences (with constant power) for 
intelligent spectrum utilization using the custom built {\em cognitive Multiple Input  Multiple  Output  (MIMO) radar} prototype. The proposed transmit waveforms avoid the frequency bands occupied by narrowband interferers or communication links, while simultaneously  have a small cross-correlation among each other to enable their separability at the MIMO radar receiver. The performance of the optimized set of sequences obtained through solving a non-convex bi-objective optimization problem, is compared with the state-of-the-art counterparts, and its applicability is illustrated by the developed prototype. 
A realistic Long Term Evolution (LTE)  downlink is used for the communications, and the real-time system implementation is validated and evaluated through the throughput calculations for communications and the  detection performance measurement for the radar system.
\end{abstract}


\IEEEpeerreviewmaketitle

\section{Introduction}

\IEEEPARstart{S}{pectrum}
congestion has become an imminent problem with multitude of radio services like wireless communications, active \gls{RF} sensing and radio astronomy vying for the scarce usable spectrum.  
Within this conundrum of spectrum congestion, radars  need to cope with simultaneous transmissions from other \gls{RF} systems. Spectrum sharing with communications being the highly plausible scenario given the need for high bandwidth in both systems \cite{8828016,8828030,Dokhanchi2019AmmWave}. While elaborate allocation policies are in place to regulate the spectral usage,  the rigid allocations result in  inefficient spectrum utilization when the subscription is sparse. In this context, smart spectrum utilization offers a flexible and a fairly promising solution for improved system performance in the emerging smart sensing systems \cite{6967722}. 

\begin{figure}
	\centering
	\includegraphics[width=1.0\columnwidth]{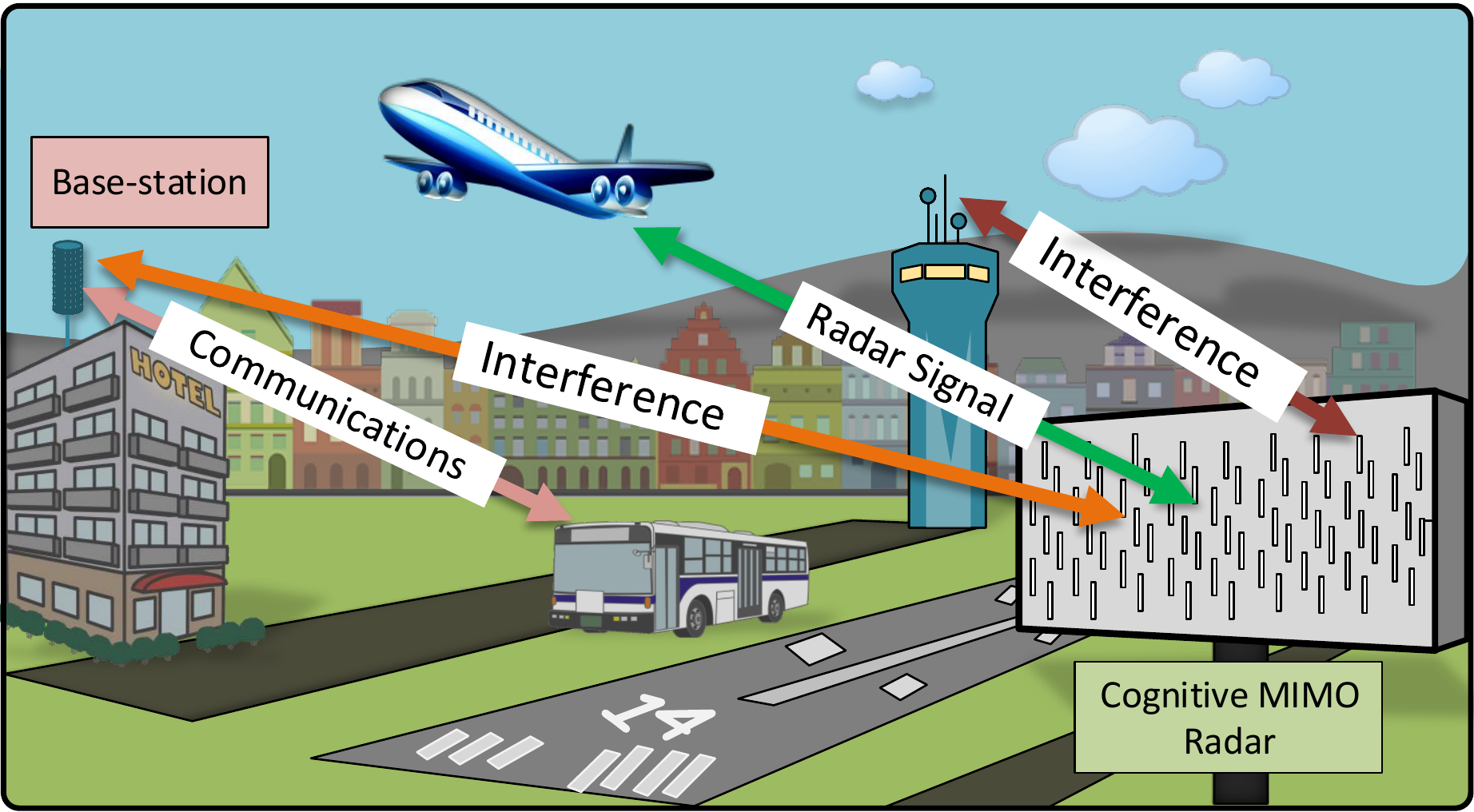}
	\caption{An illustration of coexistence between  radar and communications. The radar aims to detect the airplane, without creating interference to the communication links, and similarly avoiding interference from the communication links.}
	\label{fig:Scenario}
	\vspace*{-0.2in}
\end{figure}

Two paradigms, {\em Cognition} and {\em\gls{MIMO}} have been central to the prevalence of  smart sensing systems. Herein, the former concept offers ability to choose intelligent transmission strategies based on prevailing environmental conditions and a prediction of the behavior of the emitters in the scene, in addition to the now ubiquitous receiver adaptation \cite{6875736,7746567,8961364,9266513}. The second concept, offers a canvas of transmission strategies to the cognition manager to select from; these strategies exploit waveform diversity and the available degrees of freedom \cite{li2008mimo,khawar2016mimo}. Smart sensing opens up the possibility of coexistence of radar systems with incumbent communication systems in the earlier mentioned spectrum sharing instance. A representative coexistence scenario is illustrated in \figurename{~\ref{fig:Scenario}}, where an understanding of the environment is essential for seamless operation of radar systems while opportunistically using the spectrum allocated to communication \cite{8815508,8828016,8828030,9127852}.

In this paper, we design a cognitive \gls{MIMO} radar system towards fostering coexistence with communications; it involves spectrum sensing and  transmission  strategies adapted to the sensed spectrum while accomplishing the radar tasks and without degrading the performance  of communications. Particularly, a set of transmit sequences is designed to focus the emitted energy in the bands suggested by the spectrum sensing module while limiting the out-of-band interference. The waveforms, along with the receive processing, are designed to enhance the radar detection performance.  The designed system is then demonstrated for the representative scenario of \figurename{~\ref{fig:Scenario}} using a custom built \gls{SDR} based prototype 
developed on \gls{USRP}s\footnote {USRPs are inexpensive programmable radio platforms  used in wireless communications and sensing prototyping, 
teaching and research.} from \gls{NI} \cite{ettus,ni}. These \gls{USRP}s  operate at sub-$6$ GHz frequencies with a maximum instantaneous bandwidth of $160$ MHz.
%
\vspace*{-0.08in}
\subsection{Background and Related Works}
\paragraph{Coexistence of Communication Systems} Spectral coexistence has been found to be a promising method to alleviate the spectrum scarcity problem. For example, in the 
ISM
band, several wireless standards operate on overlapping frequency bands such as the IEEE 802.11 (popularly known as Wi-Fi) and the IEEE 802.15.4 (popularly known as ZigBee) family of standards in the $2.4$ GHz ISM band. As the ISM band does not require a license to transmit, the participating standards transmit at will (frequency and power) leading to interference and a loss of throughput\cite{HAN201653,kumar2018robust,zhang2011enabling}. Similarly, in the 5 GHz ISM band, it has been speculated that there will be interference between IEEE 802.11n/ac and \gls{LTE} \cite{kumar2019wifi}. For the two aforementioned interference scenarios, several spectral coexistence methods 
like successive Interference Cancellation (SIC)\cite{8540902,zimo} for single antenna and Technology Independent Multi-Output (TIMO) \cite{timo,sumit_wimob,sumit_thesis_phd} for multi-antennas 
have been proposed. 

\paragraph{Radar-Communication Coexistence}
The interfered  bands, including those occupied by communications, are not  useful for the radar system, and traditional radars  aim to mitigate these frequencies at their receivers. To avoid energy wasted due to transmissions on these bands while pursuing coexistence applications,  research into the transmit strategy of spectrally shaped radar waveforms has  been driving coexistence studies since the last decade  \cite{1337459,5604089,6784117,GE2016360,7279172,8267374,8579200,8943325,9052442}.  In fact, it is possible to radiate the radar waveform in a smart way by using two key elements of the cognition; spectrum sensing and spectrum sharing \cite{7906008,8961364}. 
Further,   it is possible to increase the total radar bandwidth, and consequently improve the range resolution by combining several clear bands together \cite{1337459}.
%
%
\paragraph{Radar Waveform Optimization for Coexistence} To enable waveform design in coexistence scenarios, several optimization methods including alternative optimization, \gls{MM}, \gls{CD}, \gls{ADMM}, power like iterations, etc., have been developed recently for designing waveforms exhibiting deep notches on their spectrum. Herein, we provide a brief summary of the recent works and  refer the interested readers to \cite{1337459,5438390,5604089,6329156,6784117,GE2016360,6850145,7174964,7529179,8358735,8579200,8770133,9052442,9122033,9337317} and references therein for an elaborate literature on spectral shaping optimization.

{\sl Spectral/ Template Matching:} In \cite{5604089,6784117,GE2016360,7174964,7529179,9122033,9337317} spectral/template matching\footnote{Spectral matching is when the spectrum of the designed sequence is forced to be similar to a given template in the least-squares sense.} approach is pursued to shape the spectral behaviour of the transmit sequences. The approach of template matching is also used in \cite{GE2016360} for designing sequences with a specific auto-correlation function, by considering a weighted objective of the both spectral and auto-correlation functions. 
However, matching to a template in the vicinity of the spectrum edges may decrease the possibility of synthesizing deep notches in the undesired frequency bands. 

{\sl \gls{SINR} optimization:} In \cite{8579200,9052442}, \gls{SINR} enhancement is considered to design the transmit sequences, while the spectral behaviour is considered as a constraint. Particularly, in \cite{8579200} \gls{MI} is maximized based on \gls{MM} approach while, in \cite{9052442}, a \gls{CD} based technique is utilized to maximize the \gls{SINR}; both consider a local control on the interference energy radiated in each frequency band.  Both the  \gls{SINR} and mutual information metrics require a knowledge of  the possible location of the targets, which  cannot be obtained easily in practice. Further, in case of \gls{MIMO} radar systems, maximizing \gls{MI}/ \gls{SINR} as the only design metric has the drawback of leading to a set of waveforms that are fully correlated, i. e., converting the \gls{MIMO} radar to a conventional phased-array radar \cite{4516997}. 

{\sl Stopband/ Passband levels:} In \cite{8358735,8770133}, the ratio of the maximum stopband level to the minimum passband level is considered as the objective function to shape the spectrum of the transmit waveforms with deeper notches compared to its counterparts. This metric is particularly useful in spectrally dense environments, where only the stopbands and passbands of the interfering signals can be estimated by energy detection, matched filtering or feature detection\cite{7746567}. However, \cite{8358735,8770133}, does not consider the correlation behaviour of the designed sequences towards providing the required orthogonality in \gls{MIMO} radars.

{\sl Orthogonality:} Traditionally \gls{ISL} has been used as a metric to design set of approximately orthogonal sequences in a \gls{CDM}-\gls{MIMO} radar system, i.e., having small auto- and cross-correlation sidelobes \cite{4749273,5072243,5256329,6142119,7362231,7420715,cui2017constant,9781139095174,8706639,8768085,EhsanTSP2020}. It has been shown that if a sequence set of size $M$, has atleast one $N$-length sequence, then the \gls{ISL} lower bound  
$ {\cal{B}}_{\text{ISL}} = N^2M(M - 1)$, \cite{1055219,1455966,4250296,9781139095174},   can nearly be met by a set of {unimodular} sequences generated by the algorithms proposed in \cite{4749273,5072243,5256329,6142119,7362231,7420715,cui2017constant,9781139095174,8706639,8768085,EhsanTSP2020}.

\begin{figure}
\centering
\includegraphics[width=0.85\columnwidth]{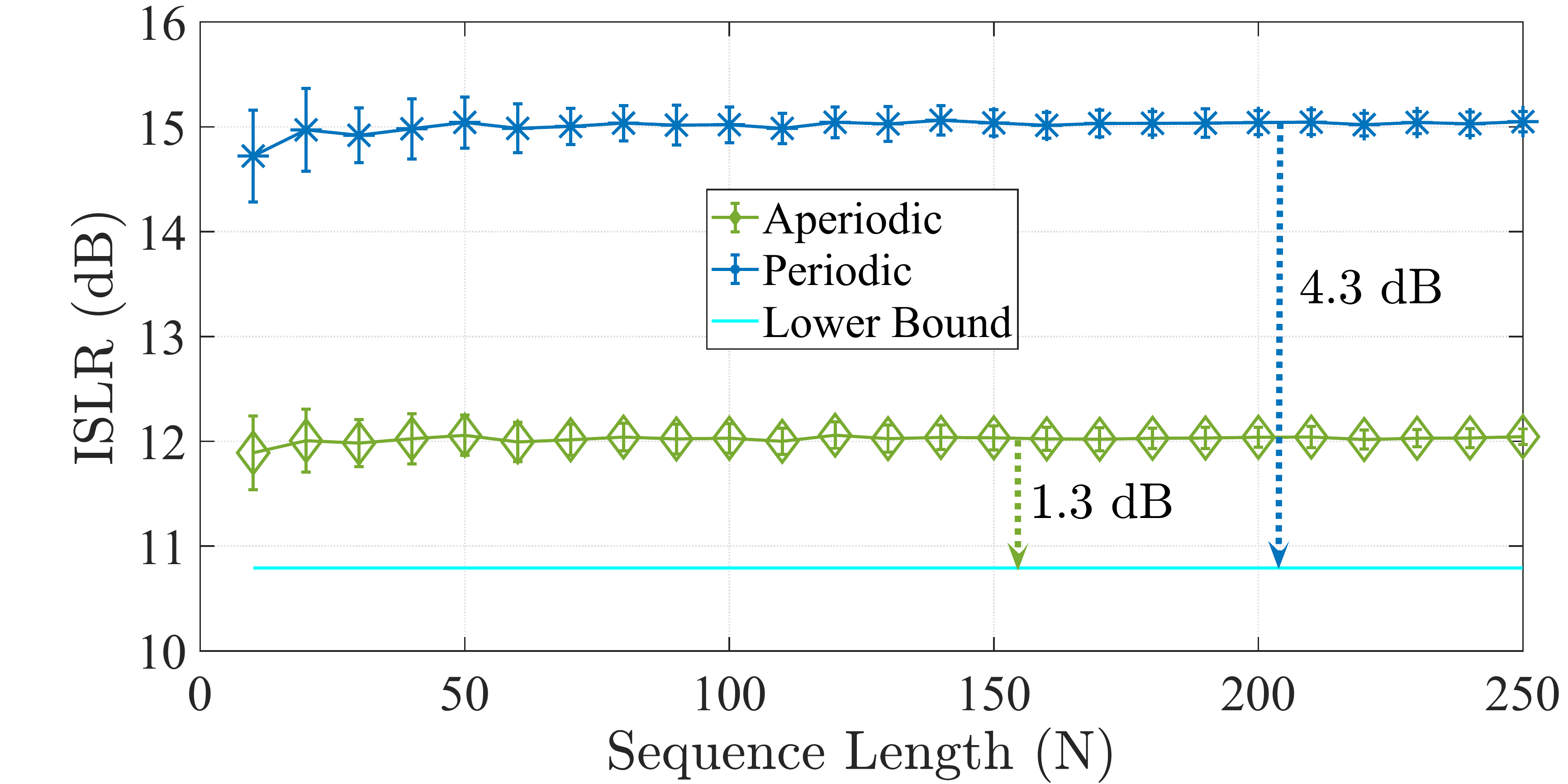}
\caption{Aperiodic/periodic \gls{ISLR} values of a set of $M=4$ random-phase sequences averaged over $100$ independent trials:  comparison for different sequence lengths.}
\label{fig:LRboundRnd}
\end{figure}

\paragraph{Summary of the Shortcomings in Prior-Art} 
The well-known notch shaping or \gls{SINR} enhancement methods do not offer satisfactory performance in terms of simultaneously providing sets of uncorrelated sequences and deep notches in the spectrum. In \figurename{~\ref{fig:LRboundRnd}}, we show that the \gls{ISLR} values of different sets of random-phase\footnote{Random-phase refers to the family of  unimodular sequences defined by
\begin{equation}
x_{m,n} = e^{j \phi_{m,n}}, ~~~ n = 1 , \ldots, N
\end{equation}
where $\phi_{m,n}$, $n = 1, \ldots, N$, are independent and identically distributed random variables with a uniform probability density function over $[0, 2\pi)$.} sequences nearly approach the lower bound on aperiodic \gls{ISLR}. This fact shows that the traditional \gls{ISL}  may not be an ideal choice for the objective function if the goal is to design set of sequences which provide orthogonality in a \gls{MIMO} radar system. Furthermore, there are no \gls{HW} prototypes to validate the waveform design in a coexistence context. Our work aims to address these shortcomings.
%
\vspace*{-0.1in}
\subsection{Contribution}\label{subSec:Contribution}
In this paper, following a radar-centric approach, we demonstrate for the first time, a coexistence set-up in incumbent communication bands by optimizing, implementing and validating a complete cognitive \gls{MIMO} radar prototype. 
The key contributions of this paper can be summarized as:
\begin{enumerate}
    \item Building a \gls{SDR} based cognitive \gls{MIMO} radar system capable of performing real-time spectrum sensing  and smart spectrum utilization for coexistence in addition to the classical radar tasks like matched filtering and Doppler processing. The testbed implements two-channels achieving a transmitting bandwidth close to $80$ MHz with the possibility of changing the parameters of the processing units during the operation.
	%
	\item Setting up of the coexistence system using the aforementioned cognitive \gls{MIMO} radar and an incumbent \gls{LTE} link which serves as the communication counterpart.
	\item  Designing a set of optimized sequences with deep notches in their spectrum that 
	can also be well-separated at the \gls{MIMO} radar receiver.  In this context, we propose an objective function based on the weighted sum  \gls{SILR} and \gls{ICCL}. These two objectives have been proposed in this paper to encourage the design of waveforms having required spectral as well as cross-correlation properties simultaneously.  The optimization is undertaken using the low complexity iterative \gls{CD} methodology which offers an attractive framework in the context of dynamic spectral shaping. In fact, the \gls{SILR} renders the \gls{CD} approach attractive through a simplified single variable optimization.
	\item Evaluating and validating the performance of the designed sequences experimentally, through a real-time visualization of the detection performance of the \gls{MIMO} radar system and a measurement of the \gls{LTE} throughput. This ensures the development, implementation and validation of one of the first prototypes for co-existing radar and communications with waveform optimization.
\end{enumerate}
%
%
%
\vspace*{-0.07in}
\subsection{Paper Overview and Notation}\label{subSec:Notation}
The rest of the paper is organized as follows. Section \ref{Sec:Prototype} describes the prototype architecture and different applications developed for the demonstration of the coexistence paradigm.  Section \ref{Sec:WaveformDesign} describes the proposed optimization of the spectral and cross-correlation of the \gls{MIMO} radar  by defining an novel objective function. In Section \ref{Sec:Perfomance}, the performance of the proposed method is compared with prior-art  and the numerical experiments of the \gls{USRP} implementation of the  prototype are given in section \ref{Sec:Experiments} and section \ref{Sec:Conclusion} summarizes the paper. 
This paper uses lower-case and upper-case boldface for vectors ($\ba$) and matrices ($\bA$) respectively. The conjugate, transpose and the conjugate transpose operators are denoted by  $(.)^*$, $(.)^T$ and $(.)^H$  respectively. The Frobenius norm, $l_2$ norm, absolute value and round operator are denoted by $\norm{.}_F$, $\norm{.}_2$, $|.|$ and $\lfloor . \rceil$ respectively. For a complex number $a$, $\Re(a)$ and $\Im(a)$ denote the real and imaginary part respectively and  $j=\sqrt{-1}$. The letter $(i)$ is use as step of a procedure. Finally $\odot$ denotes the Hadamard product.
\vspace*{-0.1in}
\section{The Proposed Prototype Architecture}\label{Sec:Prototype}
\begin{figure}
    \centering
    \begin{subfigure}{.32\linewidth}
        \centering
        \includegraphics[width=1\linewidth]{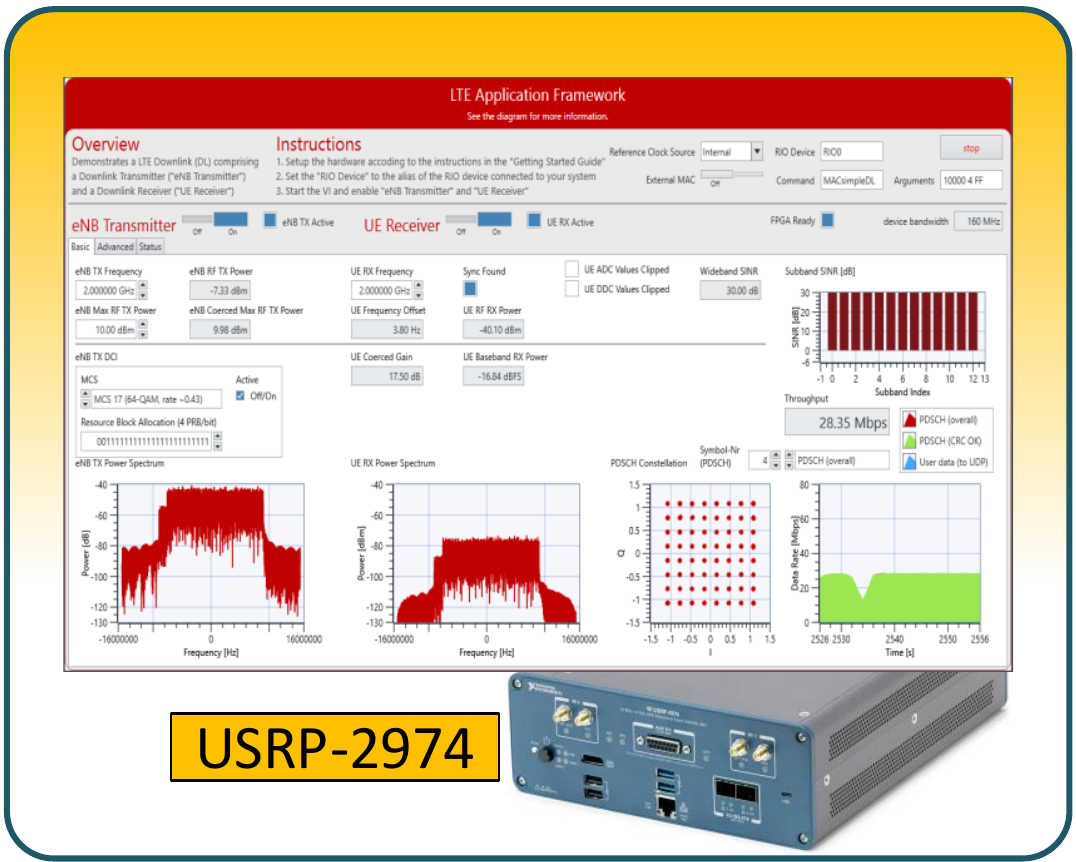}
		\caption[]{\gls{LTE} Application}\label{fig:Softwarea}
    \end{subfigure}
    \begin{subfigure}{.32\linewidth}
        \centering
        \includegraphics[width=1\linewidth]{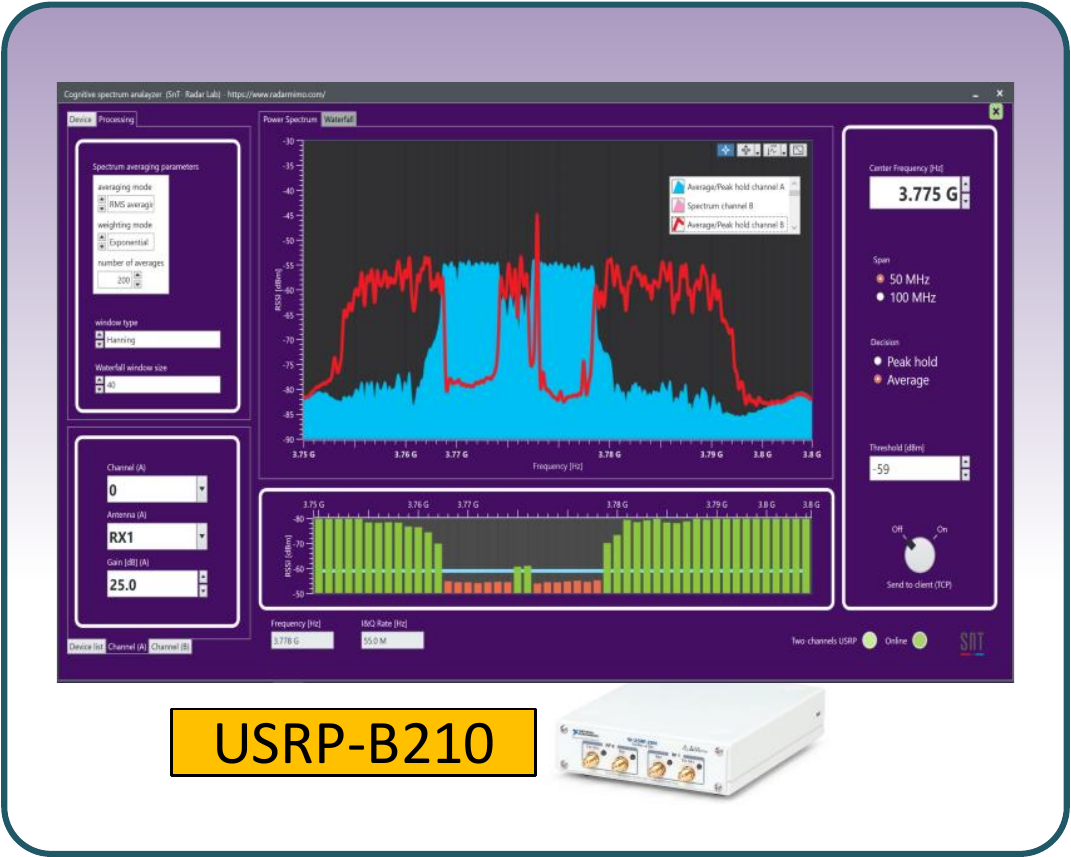}
		\caption[]{Spectrum Sensing}\label{fig:Softwareb}
    \end{subfigure}
    \begin{subfigure}{.32\linewidth}
        \centering
        \includegraphics[width=1\linewidth]{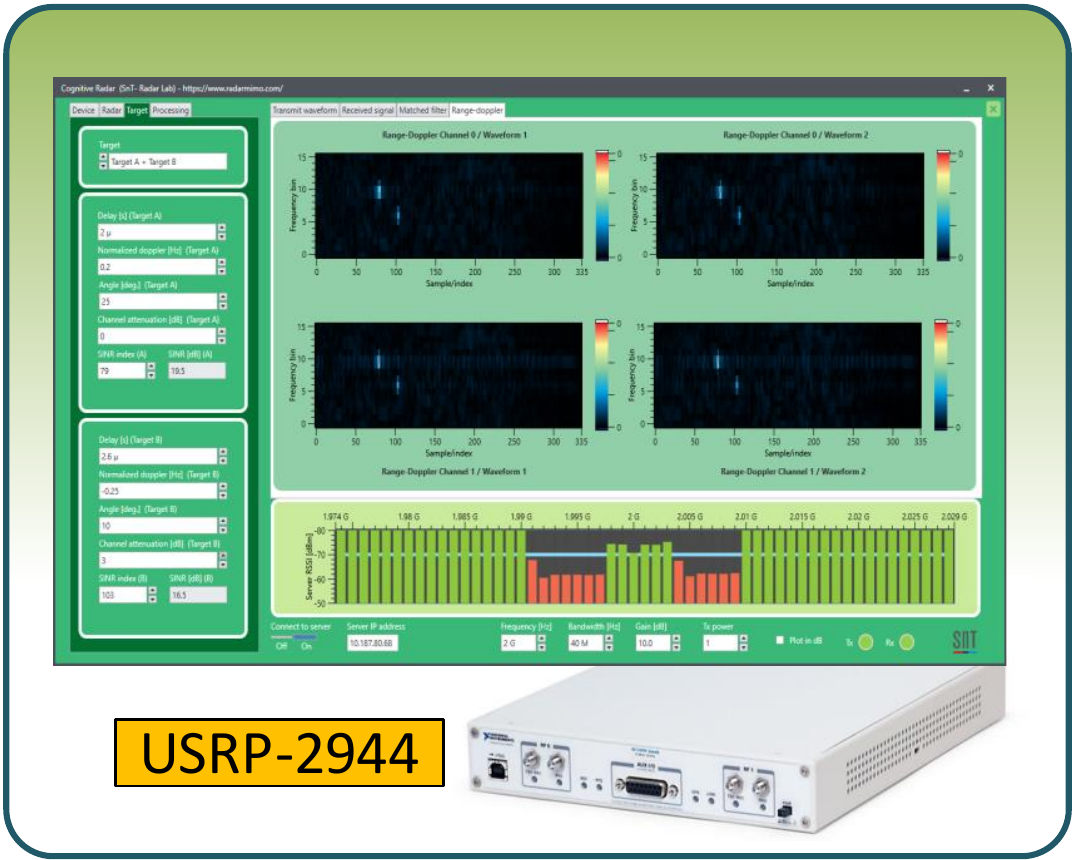}
		\caption[]{\gls{MIMO} Radar}\label{fig:Softwarec}
    \end{subfigure}
    \caption[]{Application frameworks forming the proposed prototype: \gls{LTE} application developed by NI, spectrum sensing and cognitive \gls{MIMO} radar applications developed in this paper.}\label{fig:Software}
\end{figure}

\begin{figure*}
    \centering
    \begin{subfigure}{.45\linewidth}
        \centering
        \includegraphics[width=1.0\linewidth]{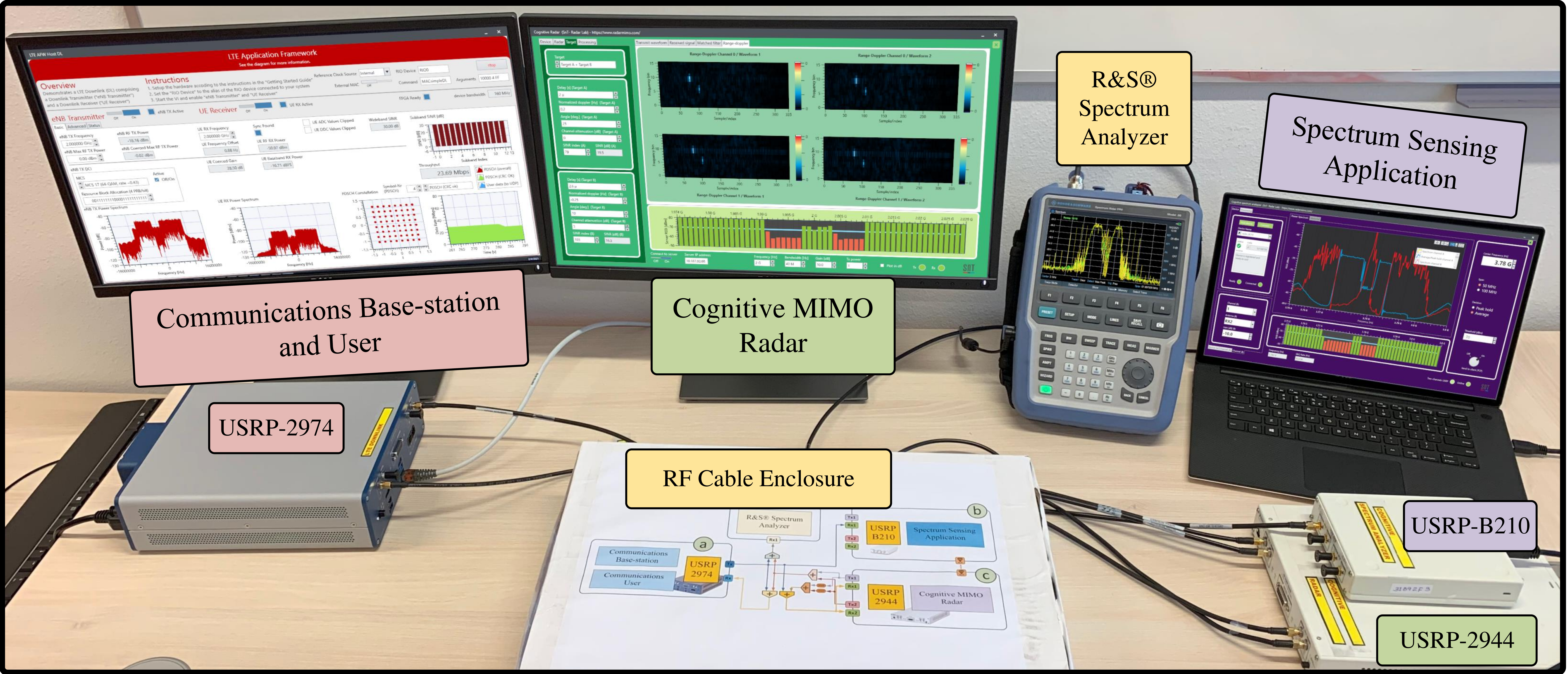}
        \caption[]{
        }
        \label{fig:SetupPhoto}
    \end{subfigure}    
    \begin{subfigure}{.45\linewidth}
        \includegraphics[width=1\linewidth]{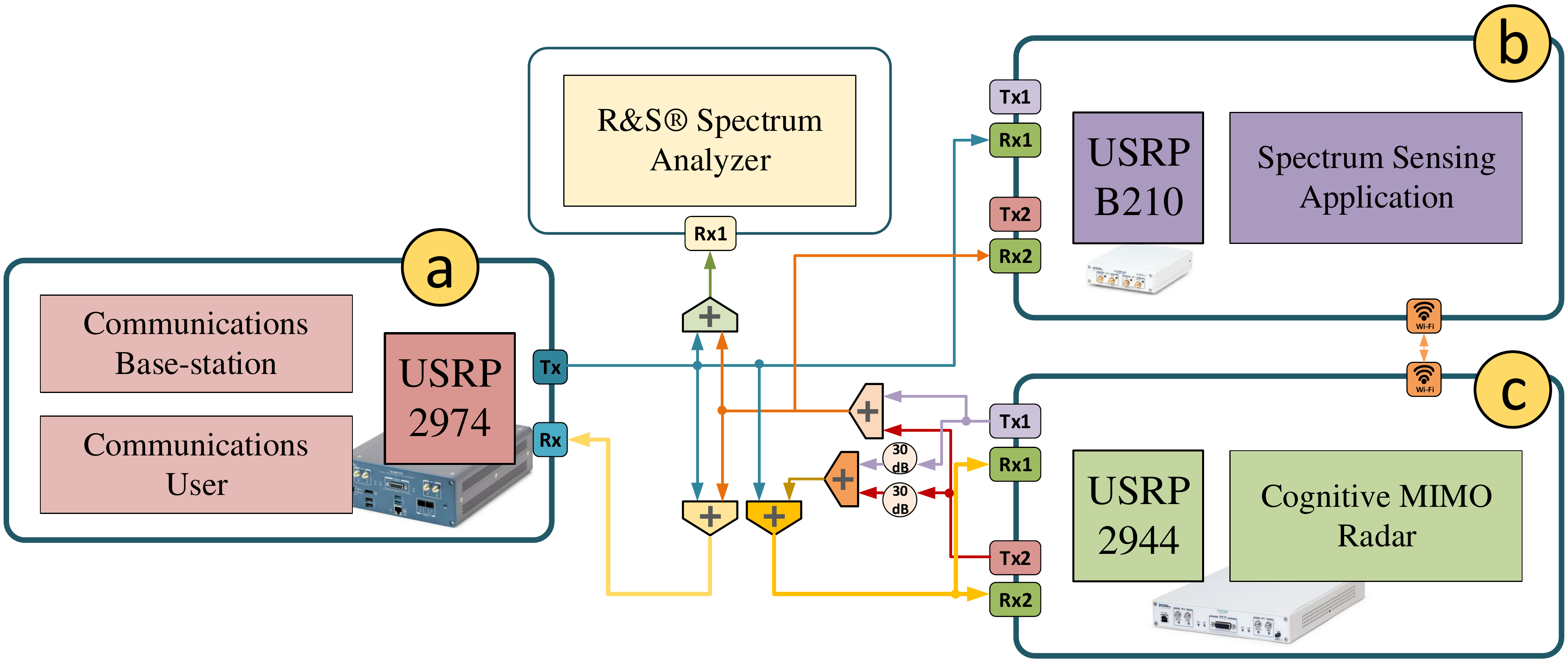}
        \caption[]{
    }
    \label{fig:BlockDiagram}
    \end{subfigure}
    \caption{ The photograph (a) and connection diagram (b) and  of the proposed coexistence prototype. The photo shows communication \gls{BS} and user, spectrum sensing, and cognitive \gls{MIMO} radar systems.
    }
\end{figure*}

The prototype consists of three application frameworks as depicted in \figurename{~\ref{fig:Software}}; a) \gls{LTE} Application Framework,
b) Spectrum sensing application, and c) Cognitive \gls{MIMO} radar application.  Various building blocks of the proposed coexistence prototype are shown in \figurename{~\ref{fig:BlockDiagram}}. The \gls{HW} consists 
of three main modules: 1) \gls{USRP} 2974 for \gls{LTE} communications
2) \gls{USRP} B210 for spectrum sensing, and 3) \gls{USRP} 2944R for cognitive \gls{MIMO} radar with specifications  given in \tablename{~\ref{tab:USRPs}}. The design details of each module are presented in the sequel. \gls{USRP}s are used for the transmission and reception of the wireless RF signals and the Rohde and Schwarz spectrum analyzer is used for the validation of the transmission\footnote{ A video of the \gls{OTA} operation of the proposed coexistence prototype can be found in  \href{ https://radarmimo.com/coexistence-of-communications-and-cognitivemimo-radar-waveform-design-and-prototype/}{radarmimo.com}. }.
%

%
\subsection{LTE Application Framework}
The LabVIEW \gls{LTE} Application Framework (\figurename{~\ref{fig:Softwarea}}) is an add-on software that provides a real-time physical layer \gls{LTE} implementation in the form of an open and modifiable source-code \cite{LTELabview}. The framework complies with a selected subset of the 3GPP \gls{LTE} which includes a closed-loop \gls{OTA} operation with channel state and ACK/ NACK feedback, $20$ MHz bandwidth, \gls{PDSCH} and \gls{PDCCH}, up to $75$ Mbps data throughput, FDD and TDD configuration $5$-frame structure, QPSK, $16$-QAM, and $64$-QAM modulation, channel estimation and zero-forcing channel equalization. 
The framework also has a basic MAC implementation to enable packet-based data transmission along with a MAC adaptation framework for rate adaptation.
Since the \gls{NI}-\gls{USRP} 2974 has two independent \gls{RF} chains and the Application Framework supports single antenna links, we emulated both the \gls{BS} and communications user on different \gls{RF} chains of the same \gls{USRP}.
\begin{table}
\centering
\caption{Hardware characteristics of the proposed prototype}
\begin{tabular}{l|l|l}
    \textbf{Parameters}                       & \textbf{2974/2944R}                     & 
    \textbf{B210}                            
    \\ 
    \hline
    Frequency range                           & $10$ MHz $ - 6$ GHz      & $70$ MHz $ - 6$ GHz              \\ 
    Max. output power                      & $20$ dBm  & $10$ dBm \\ 
    Max. input power         & $+10$ dBm    & $-15$ dBm            \\ 
    Noise figure         & $5 - 7$ dB  & $8$ dB  \\
    Bandwidth         & $160$ MHz           & $56$ MHz                   \\ 
    DACs               &  $200$ MS/s, $16$ bits   &  $61.44$ MS/s, $12$ bits    \\ 
    ADCs               & $200$ MS/s, $14$  bits   &  $61.44$ MS/s, $12$  bits   
    \\ 
\end{tabular}
\label{tab:USRPs}
\end{table}
\begin{table*}[]
\centering
\caption{Characteristics of the developed applications}
\begin{tabular}{l|l|l}
    \textbf{Parameters}                       & \textbf{Spectrum sensing}                     & 
    \textbf{Cognitive \gls{MIMO} radar}                         
    \\ 
    \hline
    \textbf{Bandwidth}                      & $50$ MHz / $100$ MHz  & $1-80$ MHz \\ 
    \hline
    \textbf{Window type}        & Rectangle, Hamming, Hanning, Blackman, etc.  &  Rectangle, Hamming, Hanning, Blackman, etc.         \\ 
    \hline
    \textbf{Averaging mode}         & RMS averaging, vector averaging, peak hold  & Coherent integration (\gls{FFT})  \\
    \hline
    \textbf{Processing units}         & 
    \begin{tabular}[c]{@{}l@{}}
    RSSI calculation in $1$ MHz bins \\
    based on spectrum waterfall average/peak
    \end{tabular} 
   & Matched filtering, range-Doppler processing             \\ 
   \hline
    \textbf{Transmitting waveforms }              & --   & 
    \begin{tabular}[c]{@{}l@{}}
    Random-polyphase, Frank,  Golomb, \\
    Random-Binary, Barker, m-Sequence, Gold, Kasami,\\
    Up-LFM, Down-LFM,\\
    and the Optimized sequences based on proposed algorithm
    \end{tabular}    \\ 
\end{tabular}
\label{tab:Apps}
\vspace*{-0.1in}
\end{table*}
\subsection{Spectrum Sensing Application}
\begin{figure*}
    \centering
    \includegraphics[width=0.7\linewidth]{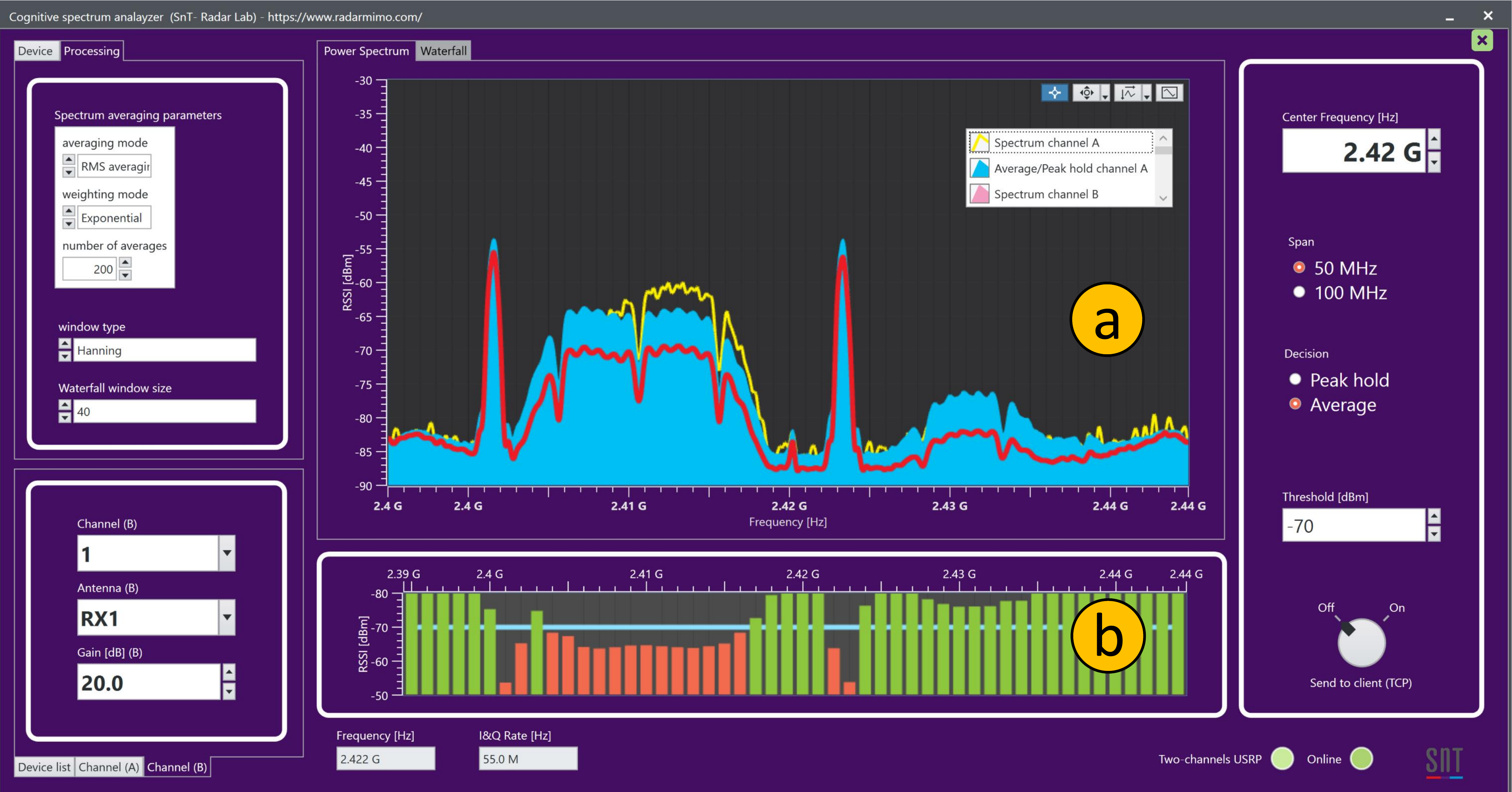}
    \caption[]{A snapshot of the developed two-channel Spectrum Sensing Application. (a) Power spectrum. (b) The energy detector. }\label{fig:SpectrumSensing}
\vspace*{-0.251in}
\end{figure*}

To perform the cognition and continuously sensing the environment, we developed an application based on LabView NXG 3.1 that connects to Ettus \gls{USRP} B2xx  (\figurename{~\ref{fig:Softwareb}}). 
The developed application is flexible in terms of changing many parameters on the fly, e.g.,  averaging modes, window type, energy detection threshold, and the \gls{USRP} configurations (gain, channel, start frequency, etc.). The features and flexibilities of the developed application are reported in \tablename{~\ref{tab:Apps}}.  \figurename{~\ref{fig:SpectrumSensing}} depicts a snapshot of the developed application when the WiFi ISM band $2.4$ GHz  with $50$ MHz bandwidth was analyzed.
The center frequency can be adjusted to any arbitrary value in the interval $70$ MHz to $6$ GHz, and the span bandwidth can be selected from the two values of $50$ MHz, and $100$ MHz\footnote{Note that \gls{USRP} B2xx provides 
$56$ MHz of real-time bandwidth by using AD9361 RFIC direct-conversion transceiver. However, the developed application can analyze larger bandwidths by sweeping the spectrum with efficient implementation.}. 
The obtained frequency chart in \figurename{~\ref{fig:SpectrumSensing}}-(b) is being transferred through a network connection (LAN/Wi-Fi) to the cognitive \gls{MIMO} radar application.

%
\vspace*{-0.1in}
\subsection{Cognitive MIMO Radar} 
The cognitive \gls{MIMO} radar application was developed based on LabView NXG 3.1, and was connected to the \gls{HW} platform NI-USRP 2944R.
This \gls{USRP} consists of a $2\times2$ \gls{MIMO} RF transceiver with a  programmable Kintex-7 field programmable gate array (FPGA). The developed application is quite  flexible in terms of changing the transmit waveform on the fly, such that it can adapt with the environment. 
\tablename{~\ref{tab:Apps}} details the features and flexibilities of the developed application. 
The center frequency can be adjusted to any arbitrary value in the interval $70$ MHz to $6$ GHz, and the radar bandwidth can be adjusted to any arbitrary value in the interval $1$ MHz to $80$ MHz.

Further, up to two targets can be also emulated in this application.  Passing the transmitting waveforms through the $30$ dB attenuators as indicated in \figurename{~\ref{fig:BlockDiagram}}, a reflection will be generated; this will be used to generate the targets, contaminated with the communications interference. The received signal in this way will be further shifted in time, frequency and spatial direction to create the simulated targets.  
These targets will be detected after calculating the absolute values of the range-Doppler maps. 

\begin{figure*}
    \centering
    \begin{subfigure}{.45\textwidth}
        \centering
        \includegraphics[width=0.8\linewidth]{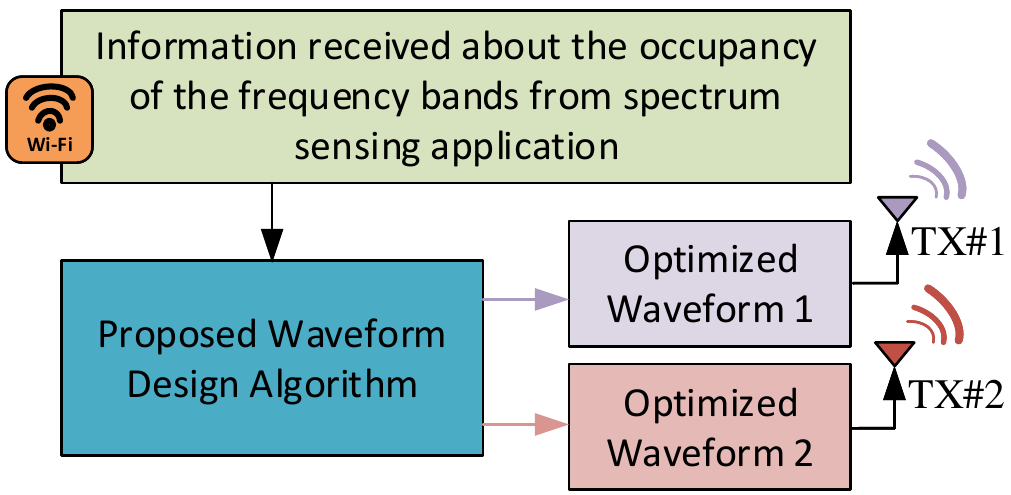}
		\caption[]{Transmitter}\label{fig:RadarBlockDiagrama}
    \end{subfigure}
    \begin{subfigure}{.45\textwidth}
        \centering
        \includegraphics[width=1\linewidth]{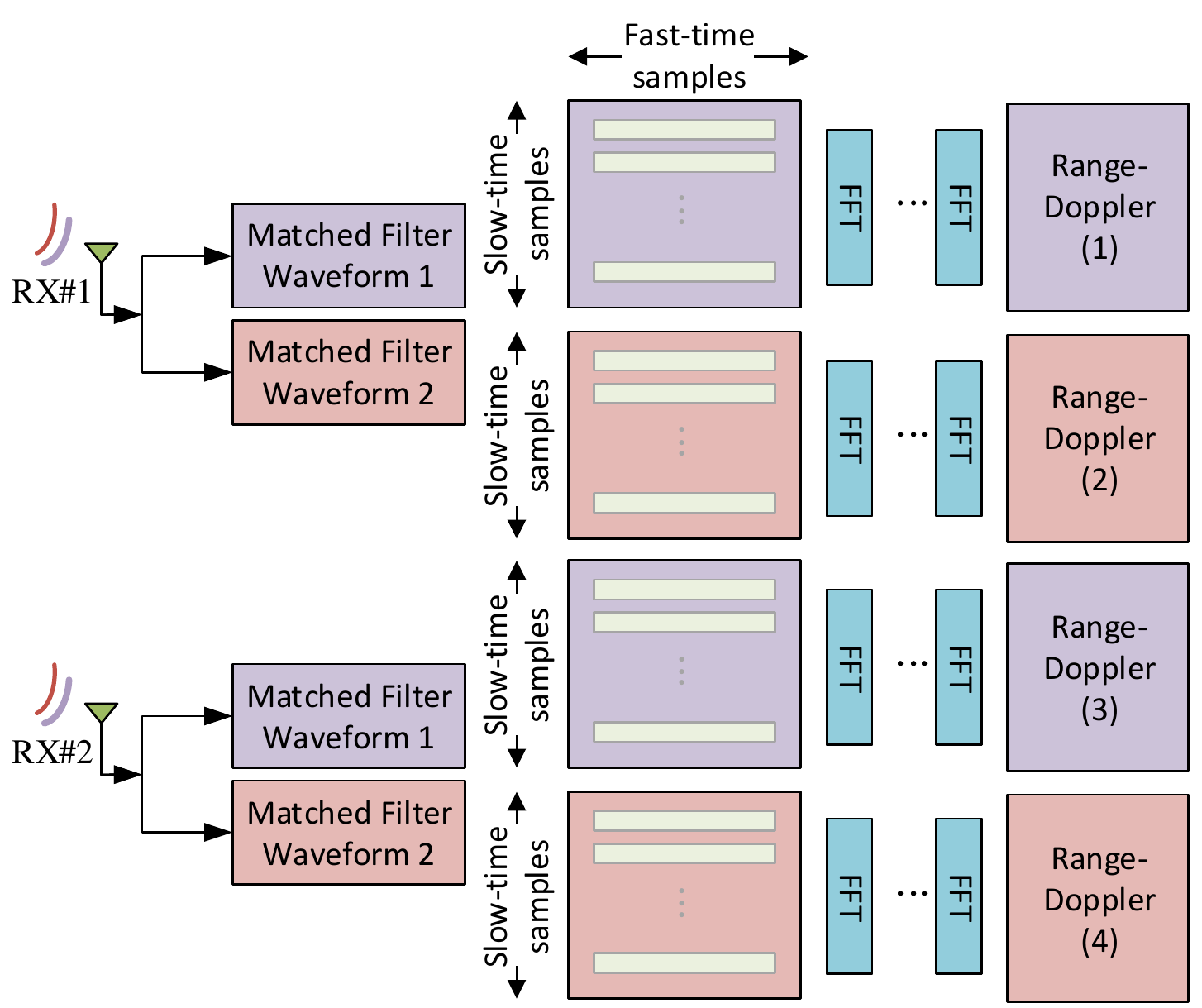}
		\caption[]{Receiver}\label{fig:RadarBlockDiagramb}
    \end{subfigure}
    \caption[]{Block diagram of the developed cognitive \gls{MIMO} radar application. In the transmit side (a), a list of occupied frequency bands will be received through the network connection between the spectrum sensing application and the radar application. Based on this information, the proposed design algorithm optimizes the transmitting waveforms. At the receive side (b), the optimized waveforms will be used for appropriate matched filtering in the fast-time dimension. Consequently, the modulus of the range-Doppler plots will be calculated after taking \gls{FFT} in the slow-time dimension.  }
    \label{fig:RadarBlockDiagram}
\end{figure*}

The block diagram of the developed cognitive \gls{MIMO} radar is depicted in \figurename{~\ref{fig:RadarBlockDiagram}}. Note that the application is connected through a network (LAN/Wi-Fi)  to the spectrum sensing application to receive the information about the occupied frequency bands. The radar adapts its transmit waveform and the matched filter coefficients based on this information. Then, using the two transmit channels of \gls{NI}-\gls{USRP} 2944R, the optimized \gls{MIMO} waveforms will be transmitted; the design algorithm for the optimization is described in Section \ref{Sec:WaveformDesign}. 

\begin{figure*}
    \centering
    \includegraphics[width=1.0\linewidth]{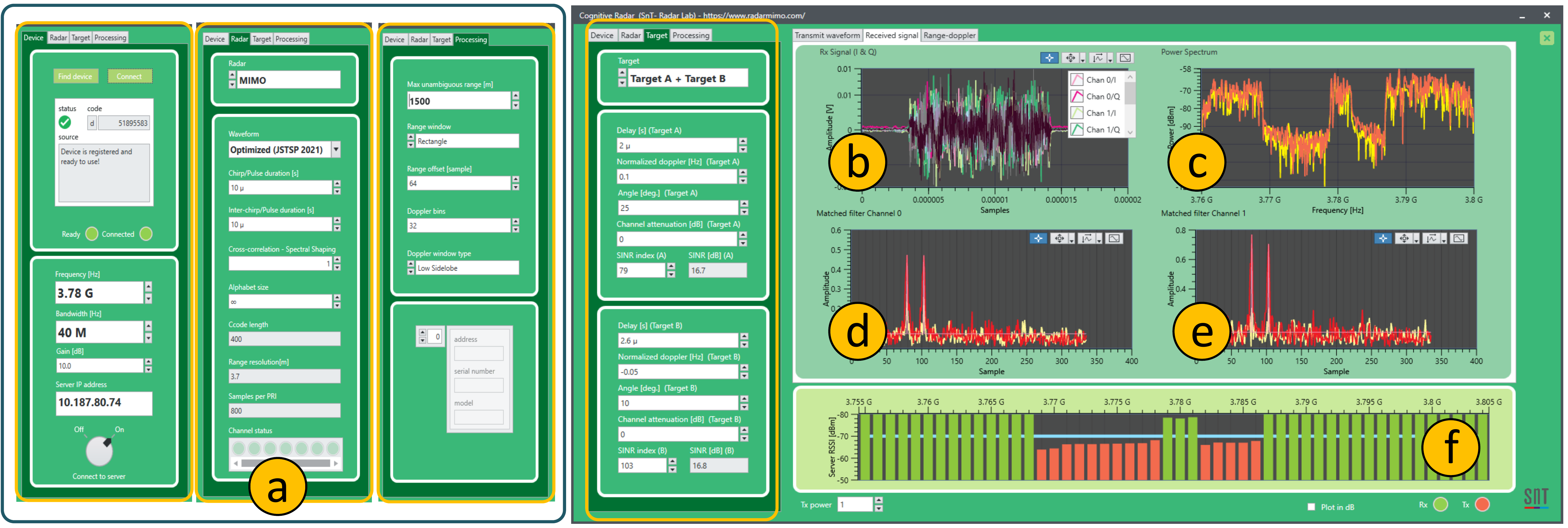}
    \caption[]{A snapshot of the developed cognitive \gls{MIMO} radar application. (a) Settings for device, radar, and processing parameters. (b) I and Q signals of two receive channels. (c) Spectrum of the received signals in two receive channels. (d) Matched filters to two transmitting waveforms at the first receive channel. (e) Matched filters to two transmitting waveforms at the second receive channel. (f) Received information from the energy detector of the spectrum sensing application. 
    }\label{fig:CognitiveMIMOradar}
\end{figure*}

\figurename{~\ref{fig:CognitiveMIMOradar}} depicts a snapshot of the developed cognitive \gls{MIMO}  radar application framework, when the licensed band $3.78$ GHz with $40$ MHz bandwidth was used for transmission\footnote{SnT has experimental licence to use $3.75$ - $3.8$ GHz for 5G research in Luxembourg.}. All the parameters related to the radar waveform, processing units, and targets can be changed and adjusted during the operation of the radar system. 

%
\section{Waveform Design}\label{Sec:WaveformDesign}
We consider a colocated narrow-band \gls{MIMO} radar system, with $M$ transmit antennas, each transmitting a sequence of length $N$ in the fast-time domain. Let the matrix $\bX \in \mathbb{C}^{M \times N} \triangleq [\mathbf{x}_1^T, \dots, \mathbf{x}_M^T]^T$ denotes the transmitted set of sequences in baseband, where the vector $\mathbf{x}_m \triangleq [x_{m,1}, \ldots, x_{m,N}]^T \in \mathbb{C}^N$ ($m=\{1,\dots,M\}$) indicates the $N$ samples of $m^{th}$ transmitter. We aim to design a transmit set of sequences which have small cross-correlation among each others, while each of the sequences having a desired spectral behaviour. 
To this end, in the following, we introduce the \gls{SILR} and \gls{ICCL} metrics and subsequently the optimization problem to handle them.


Let $\bF \triangleq [\mathbf{f}_0,\dots,\mathbf{f}_{N-1}] \in \mathbb{C}^{N \times N} $ be the \gls{DFT} matrix, where, $\mathbf{f}_k \triangleq [1, e^{j\frac{2 \pi k}{N}}, \dots, e^{j\frac{2 \pi k(N-1)}{N}}]^T \in \mathbb{C}^N, \ k = \{0, \dots, N-1\}$. Let $\mathcal{V}$ and $\mathcal{U}$ be the desired and undesired discrete frequency bands for \gls{MIMO} radar, respectively. These two sets satisfy $\mathcal{V} \cup \mathcal{U} = \{0,\dots,N-1\}$ and $\mathcal{V} \cap \mathcal{U} = \emptyset$. We define \gls{SILR} as,  
\begin{equation}\label{eq:Spectrum_ISLR}
    g_s(\bX) \triangleq \frac{\sum_{m=1}^{M} \norm{\mathbf{f}_k^{\dagger} \mathbf{x}_m}^2 | k \in \mathcal{U}}{\sum_{m=1}^{M} \norm{\mathbf{f}_k^{\dagger} \mathbf{x}_m}^2 | k \in \mathcal{V}}
\end{equation}
which is the energy of the radar waveform interfering with other incumbent services (like communications) relative to the energy of transmission in the desired bands.
Optimizing the above objective function may shape the spectral-power of the transmitting sequence and satisfy a desired mask in the spectrum. However, in a \gls{MIMO} radar it is necessary to separate the transmitting waveforms in the receiver to investigate the waveform diversity, which ideally requires orthogonality between the transmitting sequences. To make this orthogonality feasible by \gls{CDM}, we need to transmit a set of sequences which have small cross-correlations among each other. 
The aperiodic cross-correlation\footnote{In this paper, we provide the solution to the design of sequences with good aperiodic correlation functions. However, following the same steps as indicated in \cite{8768085}, the design procedure can be extended to obtain sequences with good periodic correlation properties.} of $\mathbf{x}_m$ and $\mathbf{x}_l$ is defined as,
	$r_{m,m'}(l) = \sum_{n=1}^{N-l} x_{m,n}x_{m',n+l}^*$,
where $m \neq m' \in \{1,\dots,M\}$ are  indices of the transmit antennas and $l \in \{-N+1,\dots,N-1\}$ denotes the cross-correlation lag. We define  \gls{ICCL} as, 
\begin{equation}\label{eq:ICCL}
	\widetilde{g}_c(\bX) \triangleq \sum_{m=1}^{M}\sum_{\substack{{m'=1}\\{m'\neq m}}}^{M}\sum_{l=-N+1}^{N-1}|r_{m,m'}(l)|^2,
\end{equation}
which can be used to promote the orthogonality between the transmitting sequence. 
\subsection{Problem Formulation}
We aim to design sets of sequences that simultaneously have good properties in terms of \gls{SILR} and \gls{ICCL}, under constant modulus and discrete phase constraints. The optimization problem can be represented as,
\begin{equation}\label{eq:MOOP}
	\begin{dcases}
	\min_{\bX} 	& g_s(\bX), g_c(\bX) \\
	s.t 	    & C_1 \ \text{or} \ C_2
	\end{dcases}
\end{equation}
where
$g_c(\bX) = \frac{1}{(2MN)^2} \widetilde{g}_c(\bX)$ is the scaled version of the \gls{ICCL} as defined in \eqref{eq:ICCL}. Further, $C_1 \triangleq \{\bX \mid x_{m,n} = e^{j\phi_{m,n}}, \phi_{m,n} \in \Omega_{\infty} \}, \Omega_{\infty} = [0,2\pi)$ is the constant modulus constraint and $C_2 \triangleq \{\bX \mid x_{m,n} = e^{j\phi_{m,n}}, \phi_{m,n} \in \Omega_L \}, \Omega_L = \left\{0, \frac{2\pi}{L}, \dots, \frac{2\pi(L-1)}{L}\right\}$ is discrete phase constraint. In fact, $\Omega_L$ indicates the \gls{MPSK} with alphabet size $L$.


Problem \eqref{eq:MOOP} is a bi-objective optimization problem in which a feasible solution that minimizes the both the objective functions may not exist \cite{8706639,deb2001multi}. Scalarization is a well known technique that converts the bi-objective optimization problem to a single objective problem by replacing a weighted sum of the objective functions. Using this technique, the following Pareto-optimization problem will be obtained,
\begin{equation}\label{eq:sum_weighted}
	\mathcal{P}
	\begin{dcases}
	\min_{\bX} 	& g(\bX) \triangleq \theta g_s(\bX) + (1-\theta) g_c(\bX) \\
	s.t 	    & C_1 \ \text{or} \ C_2,
	\end{dcases}
\end{equation}
The coefficient $\theta \in \left[0,1\right]$ is a weight factor that effects trade-off between \gls{SILR} and \gls{ICCL}. In \eqref{eq:sum_weighted}, $g_s(\bX)$ is a fractional quadratic function while $g_c(\bX)$ is quartic function, both with multiple variables. 
Further, both $C_1$ and $C_2$ constraints are not an affine set, besides $C_2$ is non-continuous and non-differentiate set. Therefore, we encounter a non-convex, multi-variable and NP-hard optimization problem \cite{8706639,7967829}.

\subsection{The Proposed Method}\label{Sec:proposed}
To solve \eqref{eq:sum_weighted} directly, we utilize the \gls{CD} framework wherein the multi variable problem is solved as a sequence of single variable problems. 
The methodologies based on \gls{CD}, generally start with a feasible matrix $\bX=\bX^{(0)}$ as the initial waveform set. Then, in each iteration, the waveform set is updated entry by entry several times \cite{wright2015coordinate}. In particular, an entry of $\bX$ is considered as the only variable while others are held fixed and then the objective function is optimized with respect to this identified variable. Let us assume that $x_{t,d}$ ($t \in \{1, \dots, M\}$ and $d \in \{1, \dots, N\}$) is the only variable and the fixed entries are stored in the matrix $\bX_{-(t,d)}^{(i)}$.
The resulting single-variable objective function at $(i)^{th}$ iteration can be written as (see Appendix \ref{app:1}),
\begin{equation}\label{eq:g_x_td}
\begin{aligned}
    g(x_{t,d}, \bX_{-(t,d)}^{(i)}) =& \theta \frac{a_0 x_{t,d} + a_1 + a_2 x_{t,d}^*}{b_0 x_{t,d} + b_1 + b_2 x_{t,d}^*}\\
    +& (1-\theta) \left( c_0 x_{t,d} + c_1 + c_2 x_{t,d}^* \right)
\end{aligned}
\end{equation}
where, the coefficients $a_i$, $b_i$ and $c_i$ depend on $\bX^{(i)}_{-(t,d)}$ and are derived in Appendix \ref{app:1}. By substituting $x_{t,d} = e^{j\phi}$, $\mathcal{P}$ in \eqref{eq:sum_weighted}  the optimization problem at the $i^{th}$ iteration can be recast as a function of $\phi$\footnote{For the convenience we use $\phi$ instead of $\phi_{t,d}$ in the rest of the paper.} as follows,
\begin{footnotesize}
\begin{equation}\label{eq:P_phi}
	\begin{dcases}
	\min_{\phi} 	& \theta \frac{a_0 e^{j\phi} + a_1 + a_2 e^{-j\phi}}{b_0 e^{j\phi} + b_1 + b_2 e^{-j\phi}} + (1-\theta)\left( c_0 e^{j\phi} + c_1 + c_2 e^{-j\phi} \right)  \\
	s.t 	        & C_1 \ \text{or} \ C_2,
	\end{dcases}
\end{equation}
\end{footnotesize}
At the $i^{th}$ iteration, for $t = 1, \ldots, M$, and $d = 1, \ldots, N$, the $(t,d)^{th}$ entry of $\bX$ will be updated by solving \eqref{eq:P_phi}. After updating all the entries, a new iteration will be started, provided that the stopping criteria is not met. This procedure will continue until the objective function converges to an optimal value. A summary of the proposed method is reported in \textbf{Algorithm \ref{alg:waveform_design}}. 

\begin{algorithm}
	\caption{The proposed method for designing set of sequences that avoid reserved frequency bands, and in a same time have small cross-correlation among each other.
	}
	\label{alg:waveform_design}
	\textbf{Input:} Initial set of feasible sequences, $\bX^{(0)}$.\\
	\textbf{Initialization:} $i := 0$.\\ 
	\textbf{Optimization:} 
	\begin{enumerate}
        \item {\bf while} the stopping criteria is not met, {\bf do} 
		\item \hspace{5mm} $i := i+1$;
		\item \hspace{5mm} {\bf for} $t=1,\dots,M$ {\bf do}
		\item \hspace{10mm} {\bf for} $d=1,\dots,N$ {\bf do}
		\item \hspace{15mm} Optimize  $x_{t,d}^{(i-1)}$  and obtain $x^{\star}_{t,d}$; 
		\item \hspace{15mm} Update $x^{(i)}_{t,d} = e^{j\phi^{\star}}$;
		\item \hspace{15mm} $\bX^{(i)} = \bX^{(i)}_{-(t,d)} |_{x_{t,d}=x^{(i)}_{t,d}}$;
		\item \hspace{10mm} {\bf end for}
		\item \hspace{5mm} {\bf end for}
		\item {\bf end while}
		\end{enumerate}
	\textbf{Output:} $\bX^{\star} = \bX^{(i)}$.
\end{algorithm}
Let $\phi^{\star}$ be the optimized solution of Problem \eqref{eq:P_phi}. \textbf{Algorithm \ref{alg:waveform_design}} considers a feasible set of sequences as the initial waveforms. Then, at each single variable update, it selects $x_{t,d}^{(i-1)}$ as the variable and updates it with the optimized ${s_{t,d}^{(i)}}$, denoted by ${s_{t,d}^{\star}}$. This procedure is repeated for other entries and is undertaken until all the entries are optimized at least once. After optimizing the $M N^{th}$ entry, the algorithm examines that the waveform converges to a stationary point. If the stopping criteria is not met the algorithm repeats the aforementioned steps. 
The solution to the single variable optimization problem $\mathcal{P}^{(i)}$ can be obtained by following similar steps 
of the proposed method in \cite{EhsanTSP2020}, where the the key steps are detailed in the following.

\subsubsection{Continuous phase constraint}\label{subsec:Designing Continuous phase}
The solution under continuous phase will be obtained by finding the critical points of the objective function and selecting the one that minimizes the objective. This is omitted for brevity. 


\subsubsection{Discrete phase constraint}\label{subsec:Designing Discrete phase}
In this case, the feasible set is limited to a set of $L$ phases. Thus, the objective function  with respect to the indices of $\Omega_L$ can be written as, 
\begin{equation}\label{eq:g(l)}
    g(l) = \theta\frac{\sum_{n=0}^{2} a_n e^{-j\frac{2\pi nl}{L}} }{\sum_{n=0}^{2} b_n e^{-j\frac{2\pi nl}{L}}} + (1-\theta) e^{\frac{j2\pi l}{L}}\sum_{n=0}^{2} c_n e^{-j\frac{2\pi nl}{L}}
\end{equation}
where, $l=\{0,\dots,L-1\}$. The summation term in the numerator and denominator in Equation \eqref{eq:g(l)} is exactly the definition of $L-$point \gls{DFT} of sequences $[a_0,a_1,a_2]$ , $[b_0,b_1,b_2]$ and $[b_0,b_1,b_2]$ respectively. Therefore, $g(l)$ can be written as,
\begin{equation}\label{eq:g(l)_fft}
    g(l) = \theta \frac{\mathcal{F}_L\{a_0, a_1, a_2\}}{\mathcal{F}_L\{b_0, b_1, b_2\}} + (1-\theta) \bh \odot \mathcal{F}_L\{c_0, c_1, c_2\}.
\end{equation}
where, $\bh = [1, e^{-j\frac{2\pi}{L}}, \ldots, e^{-j\frac{2\pi(L-1)}{L}}]^T \in \mathbb{C}^{L}$ and $\mathcal{F}_L$ is $L-$point \gls{DFT} operator. The current function is  valid only for $L > 2$. According to periodic property of \gls{DFT}, for binary $g(l)$ can be written as, 
\begin{equation}\label{eq:g(l)_fft2}
    g(l) = \theta \frac{\mathcal{F}_L\{a_0 + a_2, a_1\}}{\mathcal{F}_L\{b_0 + b_2, b_1\}} + (1-\theta) \bh \odot \mathcal{F}_L\{c_0 + c_2, c_1\}.
\end{equation}
Finally  
${l^{\star}} = \arg\displaystyle{\min_{l=1,\dots,L}} \left\{g(l)\right\}$, and 
${\phi_d^{\star}} = \frac{2\pi({l^{\star}}-1)}{L}. $

\section{Performance Analysis}\label{Sec:Perfomance}
This section deals with a simulation framework to compare the performance of the proposed method in terms of \gls{SILR} and \gls{ICCL} with the counterparts. 
To terminate \textbf{Algorithm \ref{alg:waveform_design}}, 
we consider
$\Delta\bX^{(i)} \triangleq \norm{\bX^{(i)} - \bX^{(i-1)}}_F \leq \zeta$ as the stopping criterion,
and set $\zeta = 10^{-5}$ for all the ensuing examples. 
Let $\mathcal{S} = \cup_{k=1}^{K_s}(s_{k,1}, s_{k,2})$ be the $K_s$ number of normalized frequency stop bands, where $0 \leq s_{k,1} < s_{k,2} \leq 1$ and $\cap_{k=1}^{K_s}(s_{k,1}, s_{k,2}) = \emptyset$. Thus, the undesired discrete frequency band is given by, $\mathcal{U} = \cup_{k=1}^{K_s}(\lfloor N s_{k,1}\rceil, \lfloor N s_{k,2}\rceil)$. 

\begin{figure}
    \centering
    \begin{subfigure}{.24\textwidth}
        \centering
		\includegraphics[width=1\linewidth]{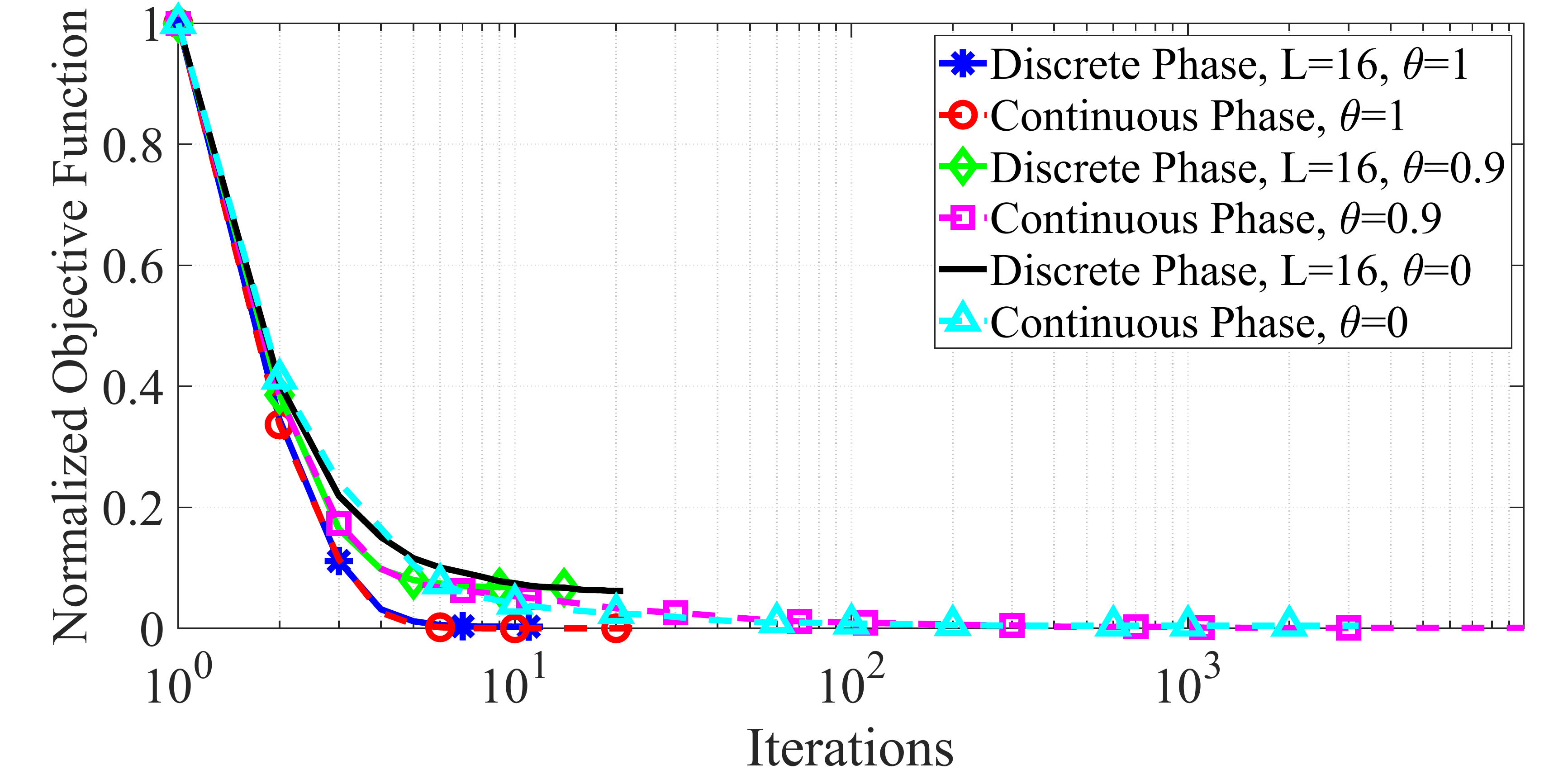}
		\caption[]{Objective function.}\label{fig:Convergence_obj}
    \end{subfigure}
    \begin{subfigure}{.24\textwidth}
        \centering
		\includegraphics[width=1\linewidth]{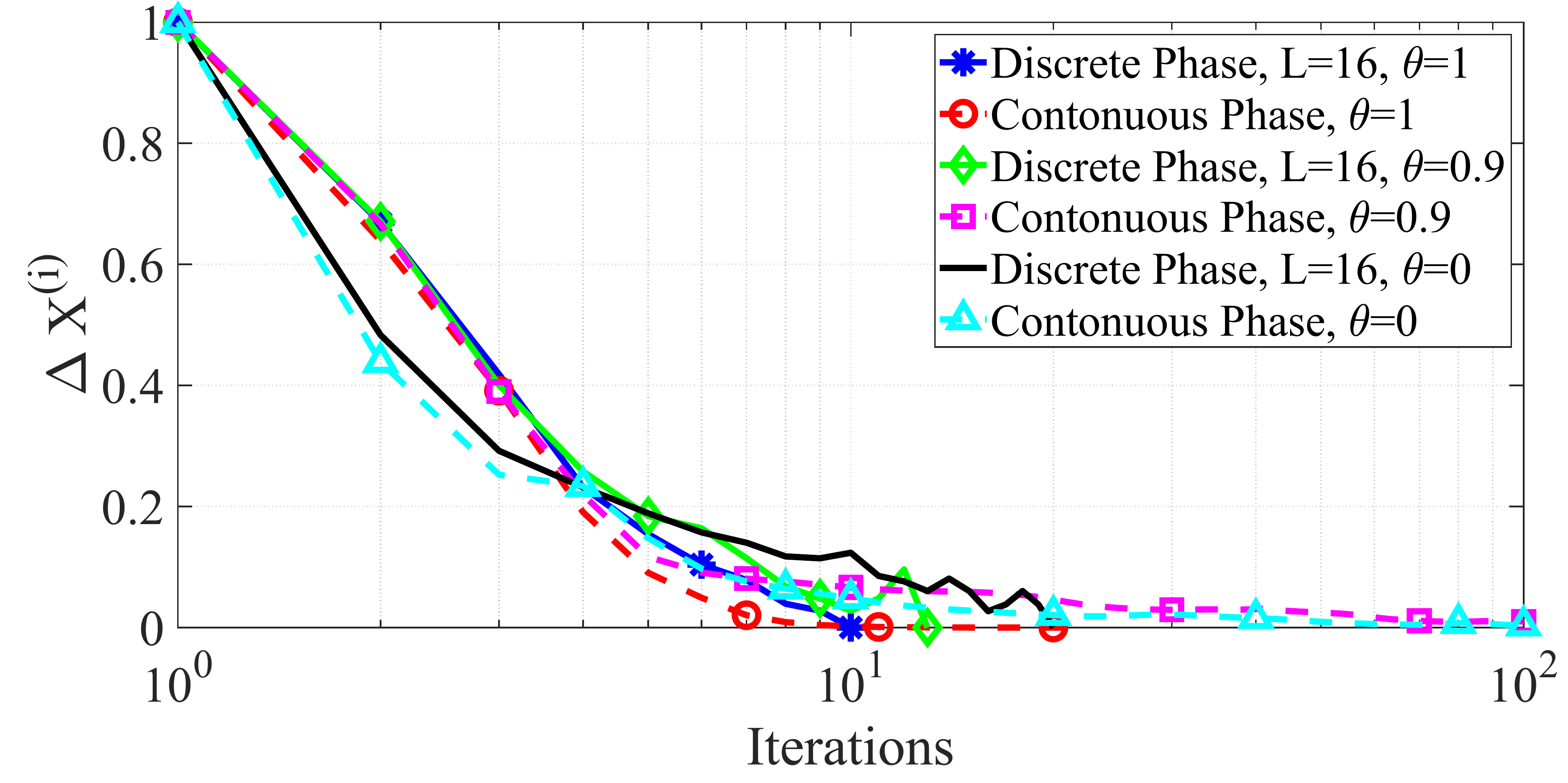}
		\caption[]{$\Delta \bX^{(i)}$.}\label{fig:Convergence_norm}
    \end{subfigure}
    \caption[]{Convergence behavior of the proposed method under continuous and discrete phase constraints for different $\theta$ values  ($M=4$, $N = 64$ and $L=16$).}\label{fig:Convergence}
\end{figure}

\figurename{~\ref{fig:Convergence_obj}} shows the convergence behavior of the objective function of the proposed method under $C_1$ and $C_2$ constraints for different values of $\theta$.  \textbf{Algorithm \ref{alg:waveform_design}} was initialized using random-phase sequences
and identical initial sequences was used for all the counterparts in case of any comparison. As can be seen, the objective function decreases monotonically and converges to a certain value under both $C_1$ and $C_2$ constraints. In case of  the $C_1$ constraint, a better performance can be observed due to the higher degrees of freedom in comparison to the $C_2$ constraint. Further, \figurename{~\ref{fig:Convergence_norm}} numerically shows the convergence of the argument to the stationary points ($\Delta \bX^{(i)}$) for different values of $\theta$. 

\begin{figure}
	\centering
	\includegraphics[width=0.9\linewidth]{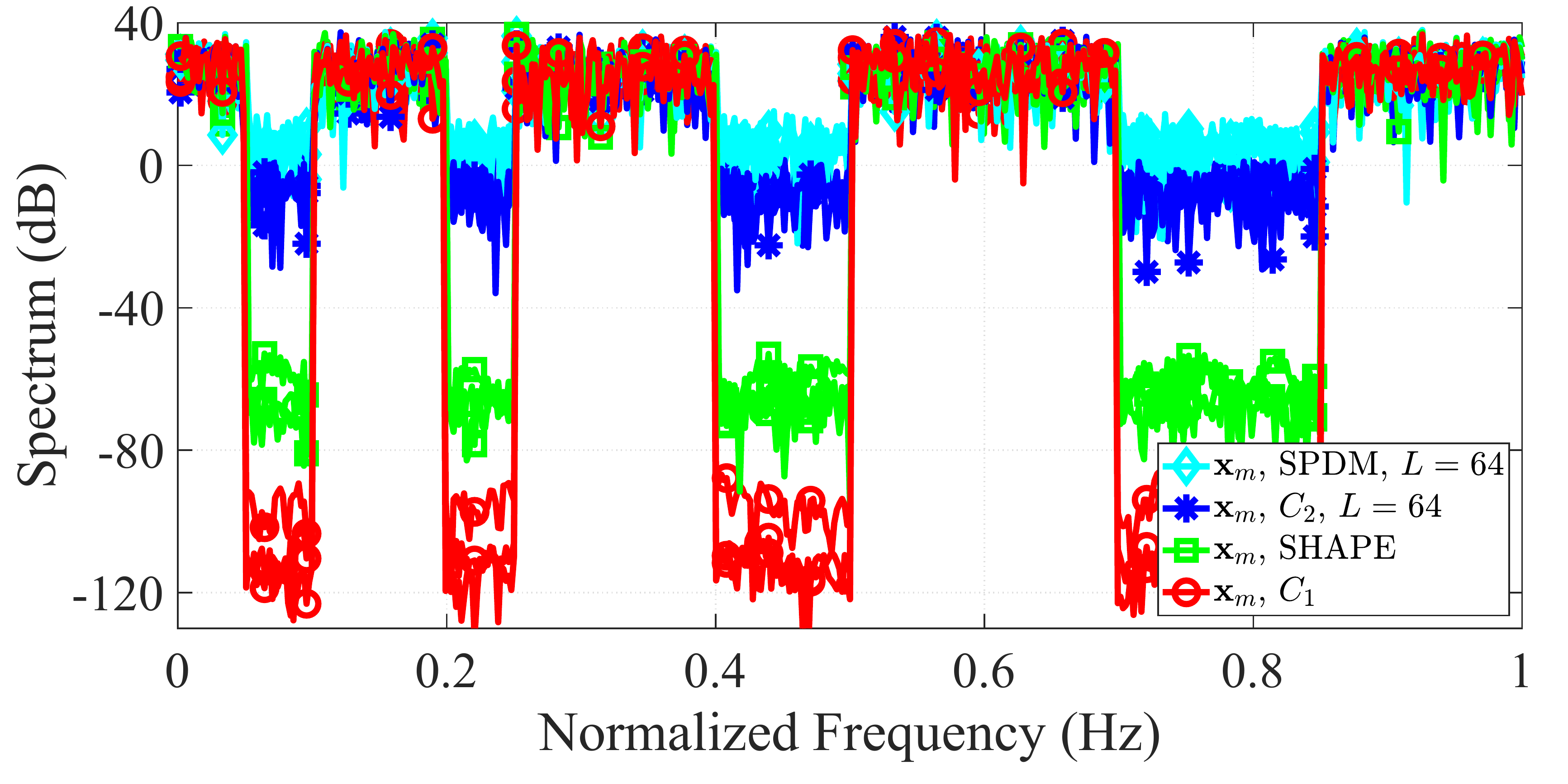}
	\caption[]{Comparing the performance of the proposed method ($\theta = 1$) with SHAPE \cite{6784117} and \gls{SDPM} \cite{9337317} in case of spectral shaping ($M=3$, $m=\{1,2,3\}$, $N = 512$, $L=64$ and $\mathcal{S} = [0.05, 0.1]$ $\cup [0.2, 0.25]$ $\cup [0.4, 0.5] $ $\cup [0.7, 0.85]$ Hz). All the methods are initialized with the same set of random-phase sequences with $L=64$.}
	\label{fig:Spectrum_Compare}
\end{figure}

In \figurename{~\ref{fig:Spectrum_Compare}}, we set $\theta = 1$ to design a set of sequences optimized based on \gls{SILR} minimization. We set $M = 3$, and denote the spectral behaviour of the optimized sequence of $m^{th}$ transmitter as $\bx_m$ in \figurename{~\ref{fig:Spectrum_Compare}}. The figure shows that the spectral behaviour of the optimized waveforms obtained by the proposed method  provide deeper notches under $C_1$ constraint, when compared with SHAPE  \cite{6784117} under a similar constraint. 
Under the $C_2$ constraint, the proposed algorithm outperforms the \gls{SDPM} \cite{9337317}, indicating the effectiveness of the proposed method  over the counterparts in shaping the spectrum of the waveforms. 

\begin{figure*}
    \centering
    \begin{subfigure}{.32\textwidth}
        \centering
		\includegraphics[width=1\linewidth]{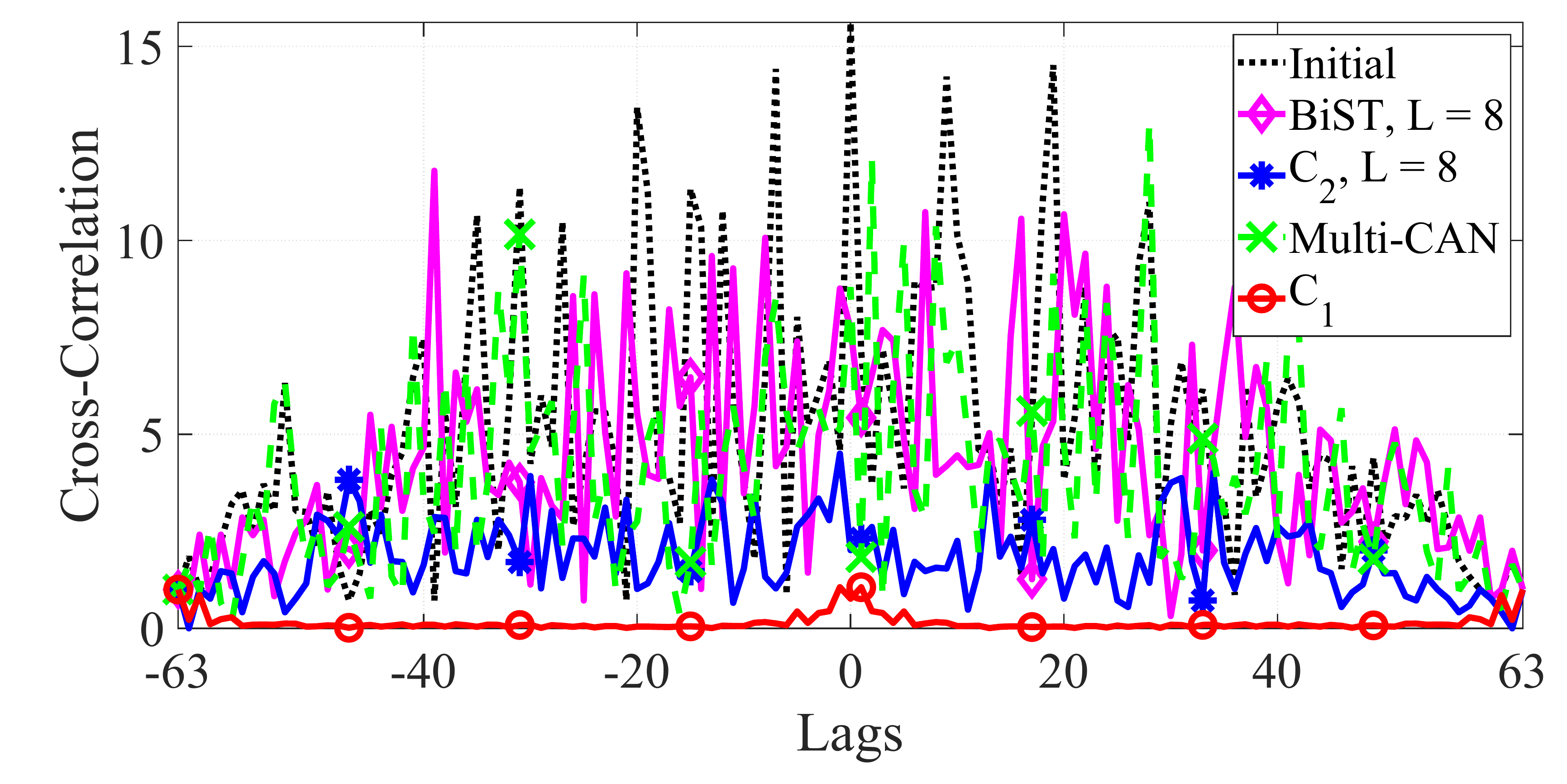}
		\caption[]{$M=3$, and $N=64$.}\label{fig:ICCL_compare}
    \end{subfigure}
    \begin{subfigure}{.32\textwidth}
        \centering
		\includegraphics[width=1\linewidth]{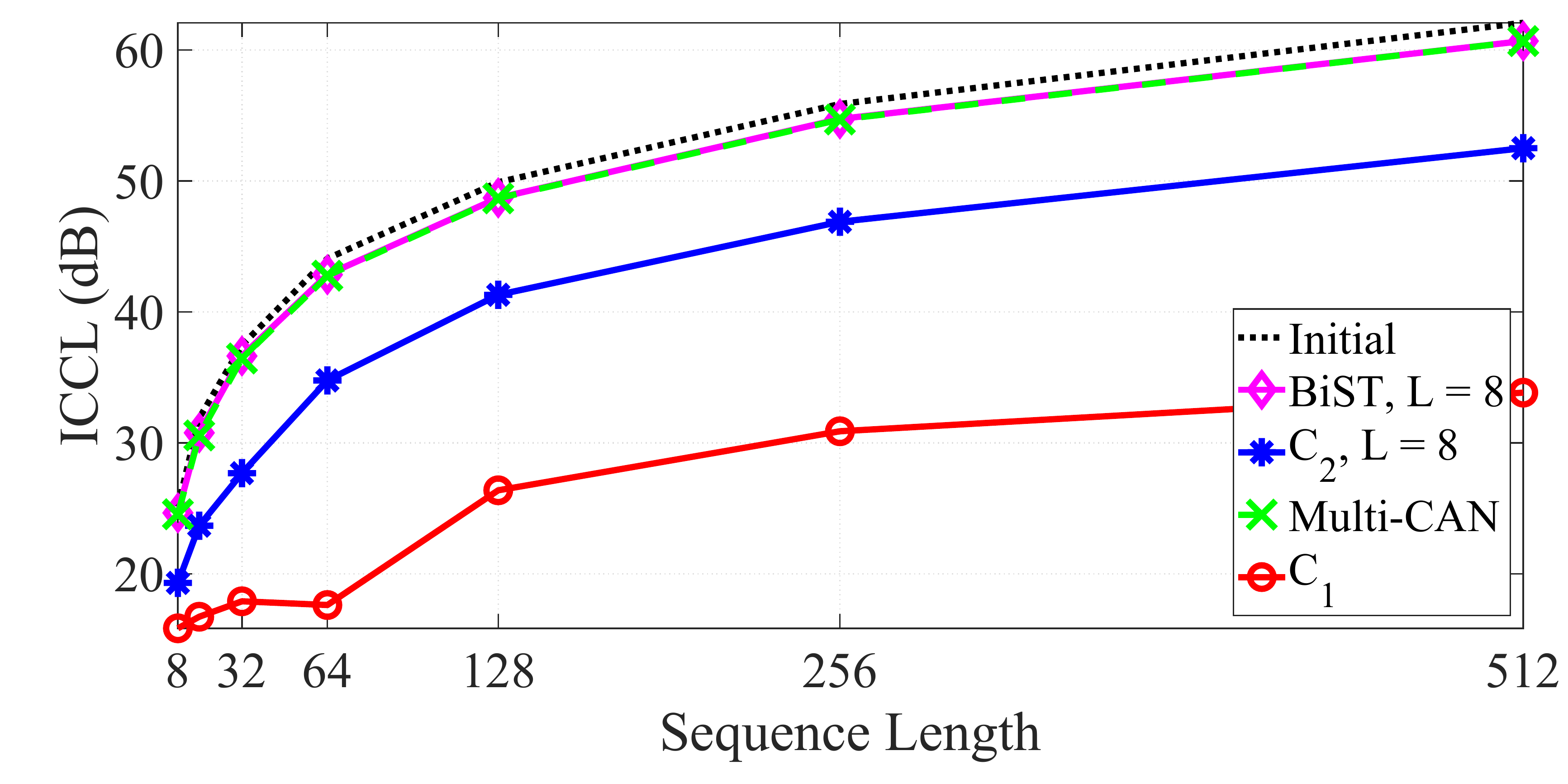}
		\caption[]{$M=3$.}\label{fig:ICCL_vs_N}
    \end{subfigure}
    \begin{subfigure}{.32\textwidth}
        \centering
		\includegraphics[width=1\linewidth]{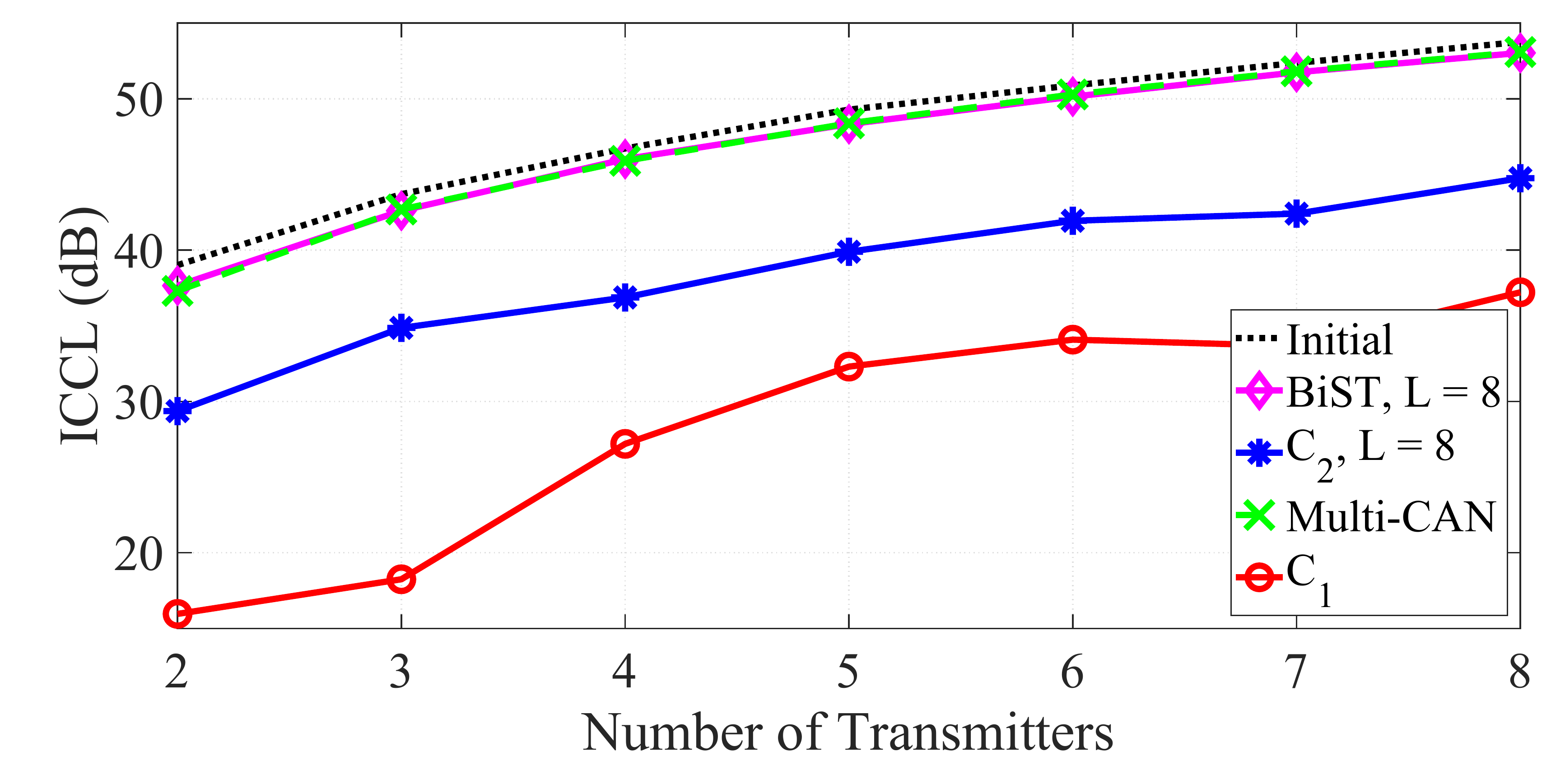}
		\caption[]{$N=64$.}\label{fig:ICCL_vs_M}
    \end{subfigure}
    \caption[]{
    Comparing the performance of the proposed method ($\theta = 0$) with Multi-CAN \cite{5072243} and \gls{BiST} \cite{8706639}  in terms of   \gls{ICCL}. All the methods are initialized with the same set of random-phase sequences. }\label{fig:ICCL}
\end{figure*}

Next, we set $\theta=0$ and evaluate the performance of the proposed method in terms of \gls{ICCL} minimization. In this case, we consider Multi-CAN \cite{5072243} and \gls{BiST} \cite{8706639} (preforming  \gls{ISL} minimization) as the benchmarks to compare the performance under $C_1$ and $C_2$ constraints. \figurename{~\ref{fig:ICCL_compare}} compares the cross-correlation of the first and second optimized waveforms obtained by the proposed method, Multi-CAN and \gls{BiST} algorithms, for $N=64$ and $M=3$. Intuitively, it can be seen that the proposed method offers set of sequences with smaller cross-correlation levels. In \figurename{~\ref{fig:ICCL_vs_N}} and \figurename{~\ref{fig:ICCL_vs_M}}, we fix $M=3$ 
and evaluate the \gls{ICCL} of the  waveforms obtained by the proposed method by changing the sequence length, and vice versa, fixing $N=64$ and changing number of transmitting waveforms. In both cases, we compare the performance of the proposed method with Multi-CAN and \gls{BiST}. It can be observed from this figure that the proposed method obtains smaller \gls{ICCL} in compare with the counterparts. 

\begin{figure}
    \centering
    \begin{subfigure}{.45\textwidth}
        \centering
		\includegraphics[width=1\linewidth]{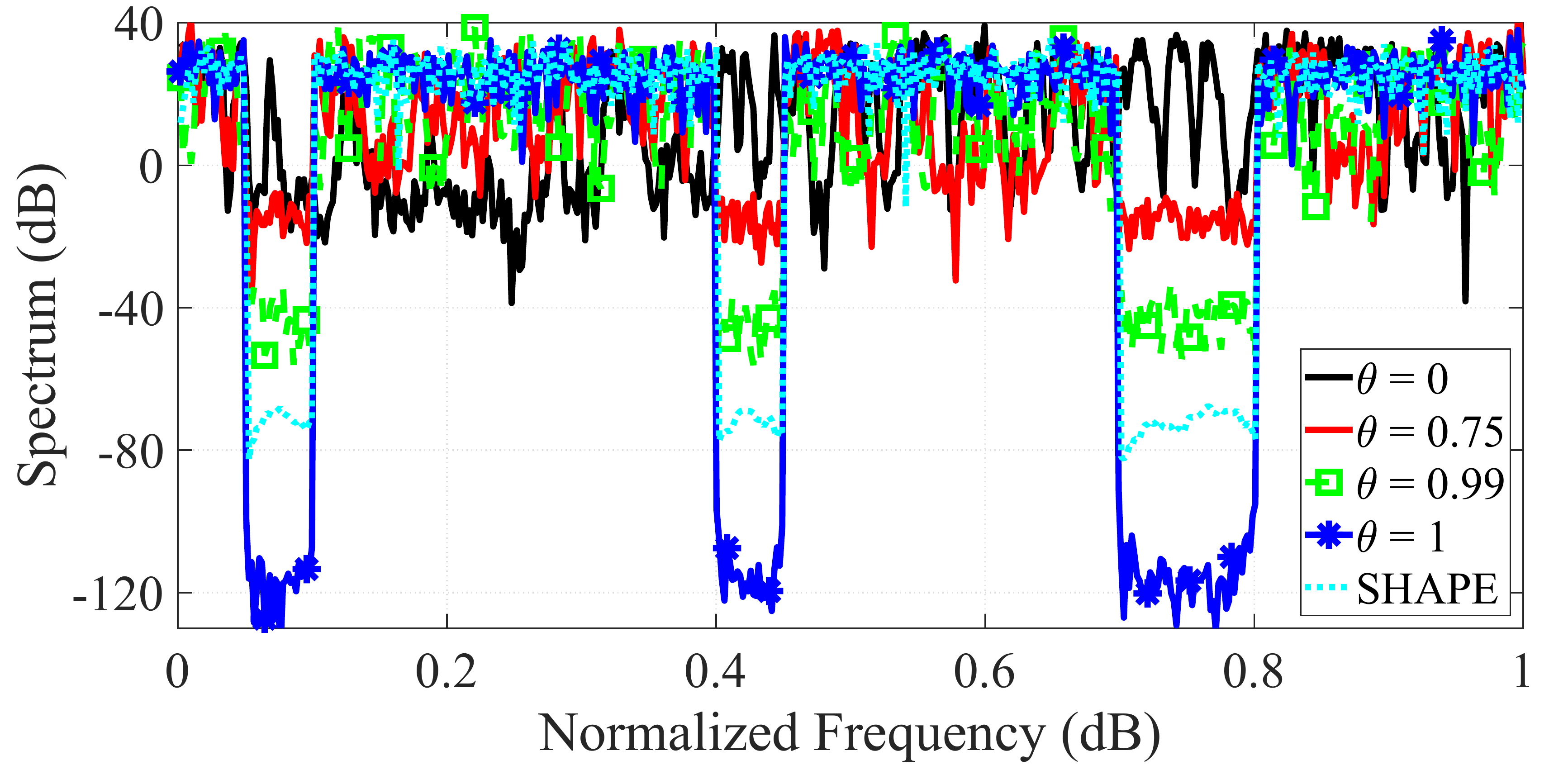}
		\caption[]{Spectrum of the proposed method for different $\theta$ values  ($\mathcal{S} = [0.05, 0.1] $ $\cup [0.4, 0.45] $ $\cup [0.7, 0.8]$ Hz).}\label{fig:Spectrum_vs_theta}
    \end{subfigure}
    \begin{subfigure}{.45\textwidth}
        \centering
		\includegraphics[width=1\linewidth]{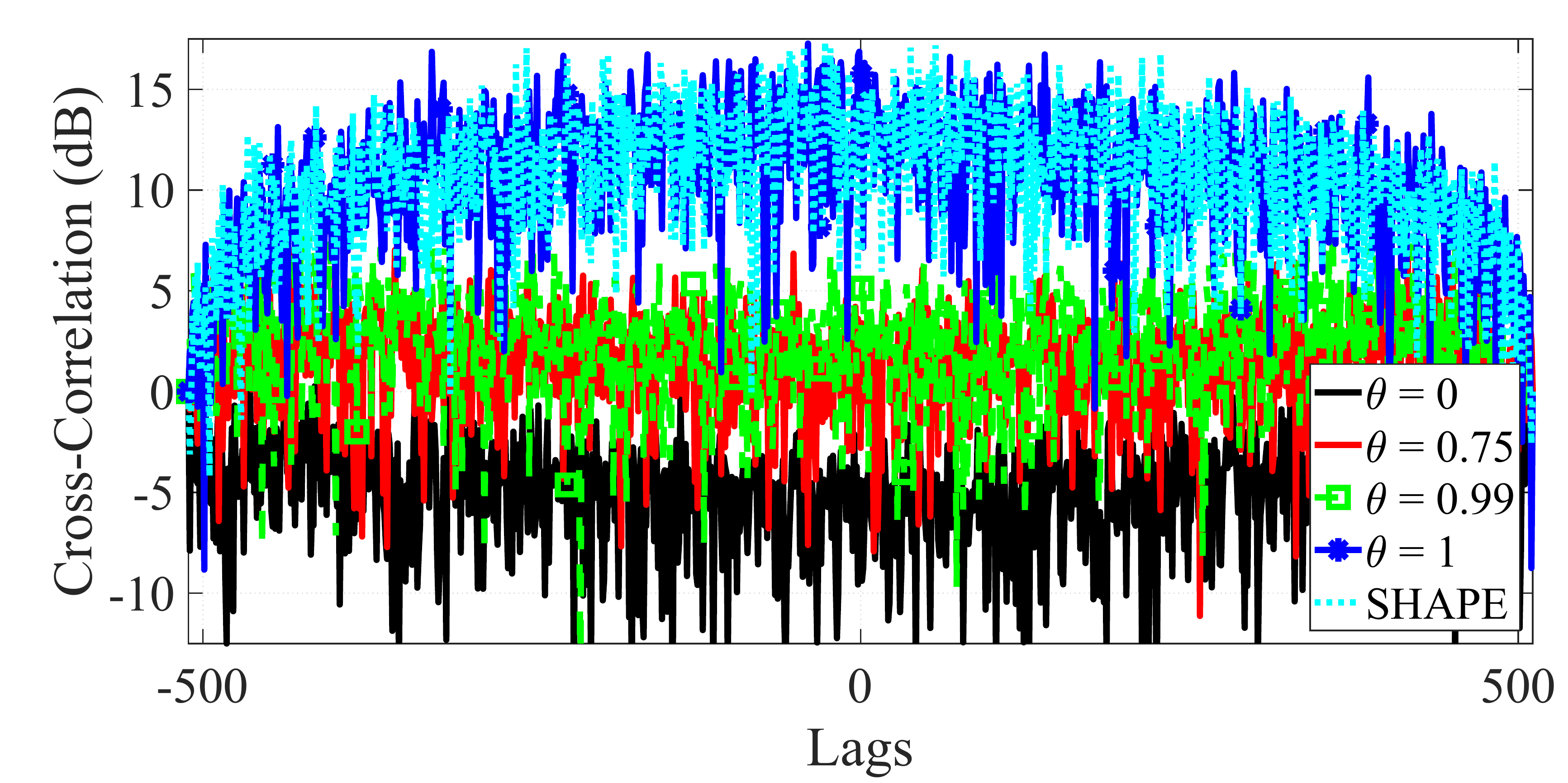}
		\caption[]{Cross-Correlation of the proposed method for different $\theta$ values.}\label{fig:Orthogonality_vs_theta}
    \end{subfigure}
    \caption[]{The impact of $\theta$ value on trade-off between (a) spectral shaping and (b) cross-correlation levels in comparision with SHAPE \cite{6784117} ($M=2$ and $N = 512$).}\label{fig:Trade-off}
\end{figure}

In \figurename{~\ref{fig:Trade-off}}, we evaluate the trade-off between a good spectrum shaping and good orthogonality by choosing $\theta$. \figurename{~\ref{fig:Spectrum_vs_theta}} shows the impact of $\theta$ on the spectral behaviour. As can be seen, by choosing $\theta = 0$, the optimized waveforms are not able to put notches on the undesired frequencies. By increasing $\theta$,  the notches will appear gradually and in case of $\theta = 1$, we obtain the deepest notches. However, when $\theta = 1$ the cross-correlation is at the highest level which decreases with $\theta$. In $\theta = 0$, we obtain the best orthogonality. This fact is shown in \figurename{~\ref{fig:Orthogonality_vs_theta}}. Therefore, by choosing an appropriate value of $\theta$, one can make a good trade-off between these two metrics. For instance, choosing $\theta = 0.75$ is able to put a null level around $50$ dB (see \figurename{~\ref{fig:Spectrum_vs_theta}}), while having good cross-correlation level (see \figurename{~\ref{fig:Orthogonality_vs_theta}}). 

\section{Experiments and Results} \label{Sec:Experiments}
In this section, we present present experiments conducted using the developed prototype and analyze the \gls{HW} results.
The prototype setup in \figurename{~\ref{fig:SetupPhoto}} depicts the cognitive \gls{MIMO} radar, the spectrum sensing application and the \gls{LTE} application framework as well as \gls{RF} cable enclosure, \gls{USRP}s, and the spectrum analyzer.
For the practical applicability of our methods and  verification of the simulation, 
we established all the connections shown in \figurename{~\ref{fig:BlockDiagram}} using \gls{RF} cables and splitters/ combiners, and 
measured the performance in a controlled environment. 
Nevertheless, a video of the \gls{OTA} operation of the proposed prototype can be found in \href{ https://radarmimo.com/coexistence-of-communications-and-cognitivemimo-radar-waveform-design-and-prototype/}{radarmimo.com}. 

The developed cognitive \gls{MIMO} radar system in the proposed coexistence prototype consists of a $2 \times 2$ transmit/receive unit. Two targets with adjustable attenuation paths, Doppler and spatial angles can be also augmented in the receive side as described in Section \ref{Sec:Prototype}. The transmitting waveforms can be selected based on the options in \tablename{~\ref{tab:Apps}} or obtained based on \textbf{Algorithm \ref{alg:waveform_design}}. 
When executing the application, 
input parameters to optimize the waveforms pass from the \gls{GUI} to MATLAB, and the optimized set of sequences 
are passed to the application through the \gls{GUI}. The other processing blocks of the radar system including matched filtering, Doppler processing, and scene generation are developed in the LabView G dataflow application.
\tablename{~\ref{tab:radar_param}} and \tablename{~\ref{tab:target_param}} summarize the parameters  used for radar and targets in this experiment. 
\begin{table}[t!]
	\centering
	\caption{Radar experiment parameters}
	\begin{tabular}{l|l}
		\hline
		\textbf{Parameters}                       & \textbf{Value}                     \\ 
		\hline
		Center frequency                           & $2$ GHz                    \\ 
		Real-time bandwidth        & $40$MHz \\
        Transmit and receive channels                           & $2 \times 2$                    \\
		Transmit power                      & $10$ dBm \\
		Duty cycle                      & $50 \% $ \\ 
		Transmit code length                      & $400$ \\
		Pulse repetition interval                      & $20 \mu $s \\ 
		\hline
	\end{tabular}
	\label{tab:radar_param}
\end{table}
\begin{table}[t!]
	\centering
	\caption{Target experiment parameters}
	\begin{tabular}{l|l|l}
		\hline
		\textbf{Parameters}                       & \textbf{Target 1} & \textbf{Target 2}                      \\ 
		\hline
		Range delay                     & $2 \mu$s   & $2.6 \mu$s  \\
		Normalized Doppler                       & $0.2$ Hz & $-0.25$ Hz\\
		Angle                     & $25 \deg$ &  $15 \deg$ \\
		Attenuation                       & $30$ dB & $35$ dB \\
		\hline
	\end{tabular}
	\label{tab:target_param}
\end{table}

For the LTE communications, we established the downlink  between  a \gls{BS} and one user. Nonetheless, the experiments can be also be performed with uplink \gls{LTE} as well as bi-directional \gls{LTE} link. 
LabVIEW \gls{LTE} framework offers the possibility to vary the \gls{MCS} of \gls{PDSCH} from $0$ to $28$ where the constellation size goes from QPSK to $64$QAM \cite{LTE_NI}. \gls{LTE} uses \gls{PDSCH} for the transport of data between the \gls{BS} and the user. \tablename{~\ref{tab:comm_param}} indicates  the  experimental  parameters  used  in our  test  set-up for the communications. 
\begin{table}[t!]
	\centering
	\caption{Communications experiment parameters}
	\begin{tabular}{l|l}
		\hline
		\textbf{Parameters}                       & \textbf{Value}                     \\ 
		\hline
		Communication \gls{MCS}         & \begin{tabular}[c]{@{}l@{}}MCS0 (QPSK $0.12$)\\ MCS10 (16QAM $0.33$) \\ MCS17 (64QAM $0.43$)\end{tabular} \\ 
		Center frequency (Tx and Rx)          & $2$ GHz \\
		Bandwidth         & $20$ MHz \\
		\hline
	\end{tabular}
	\label{tab:comm_param}
\end{table}
\subsection{Impact of SILR Minimization}
In this experiment, we  show the impact of \gls{SILR} enhancement on the coexistence.  We use the experiment parameters reported in \tablename{~\ref{tab:radar_param}}  and \tablename{~\ref{tab:comm_param}} for radar and communications, respectively. 

\begin{figure*}
\centering
    \begin{subfigure}{.31\textwidth}
        \centering
		\includegraphics[width=1\linewidth]{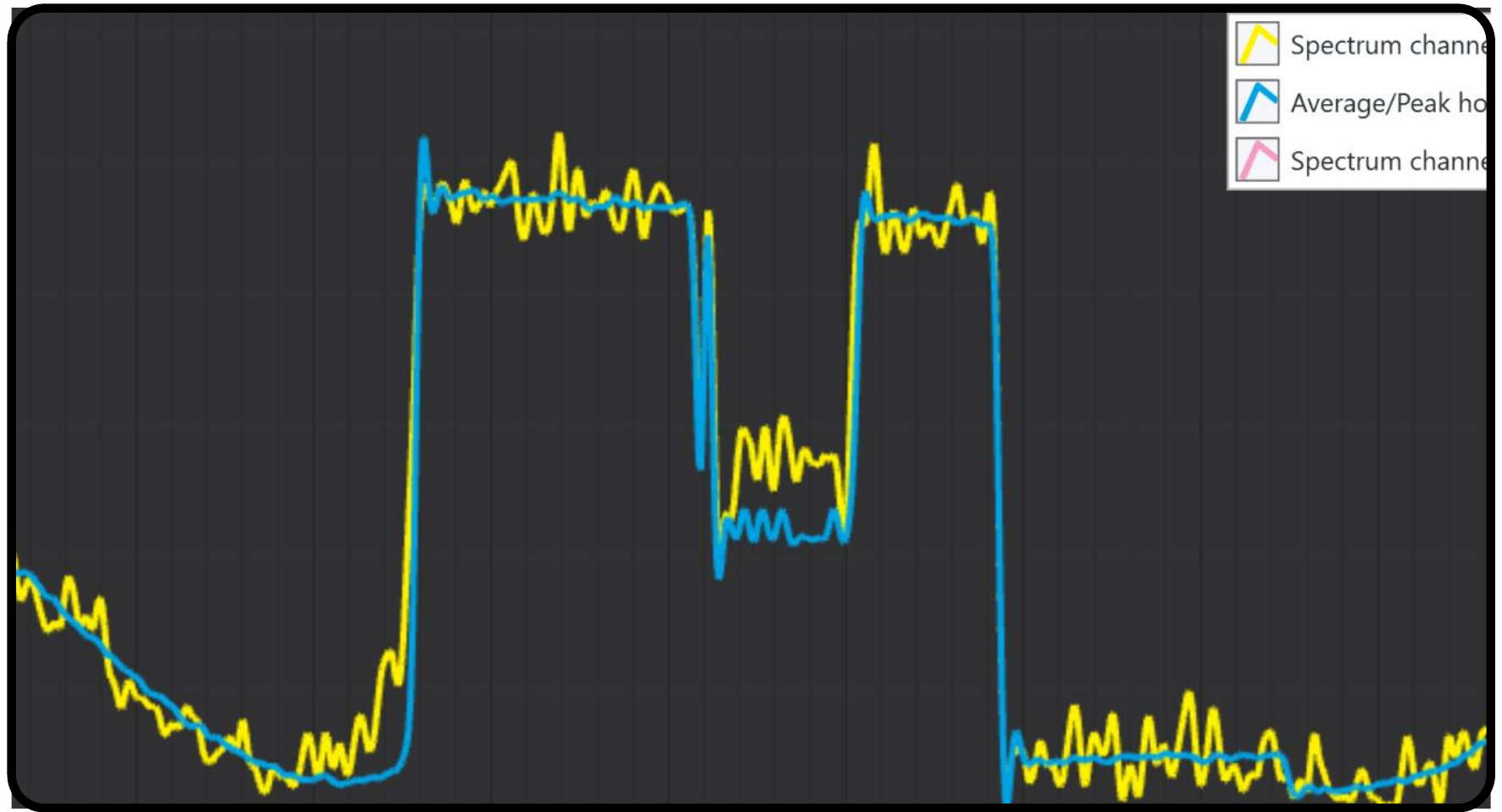}
		\caption[]{\gls{LTE} at developed application.}
		\label{fig:SpectrumMeasurment2a}
    \end{subfigure}  
    \begin{subfigure}{.31\textwidth}
        \centering
		\includegraphics[width=1\linewidth]{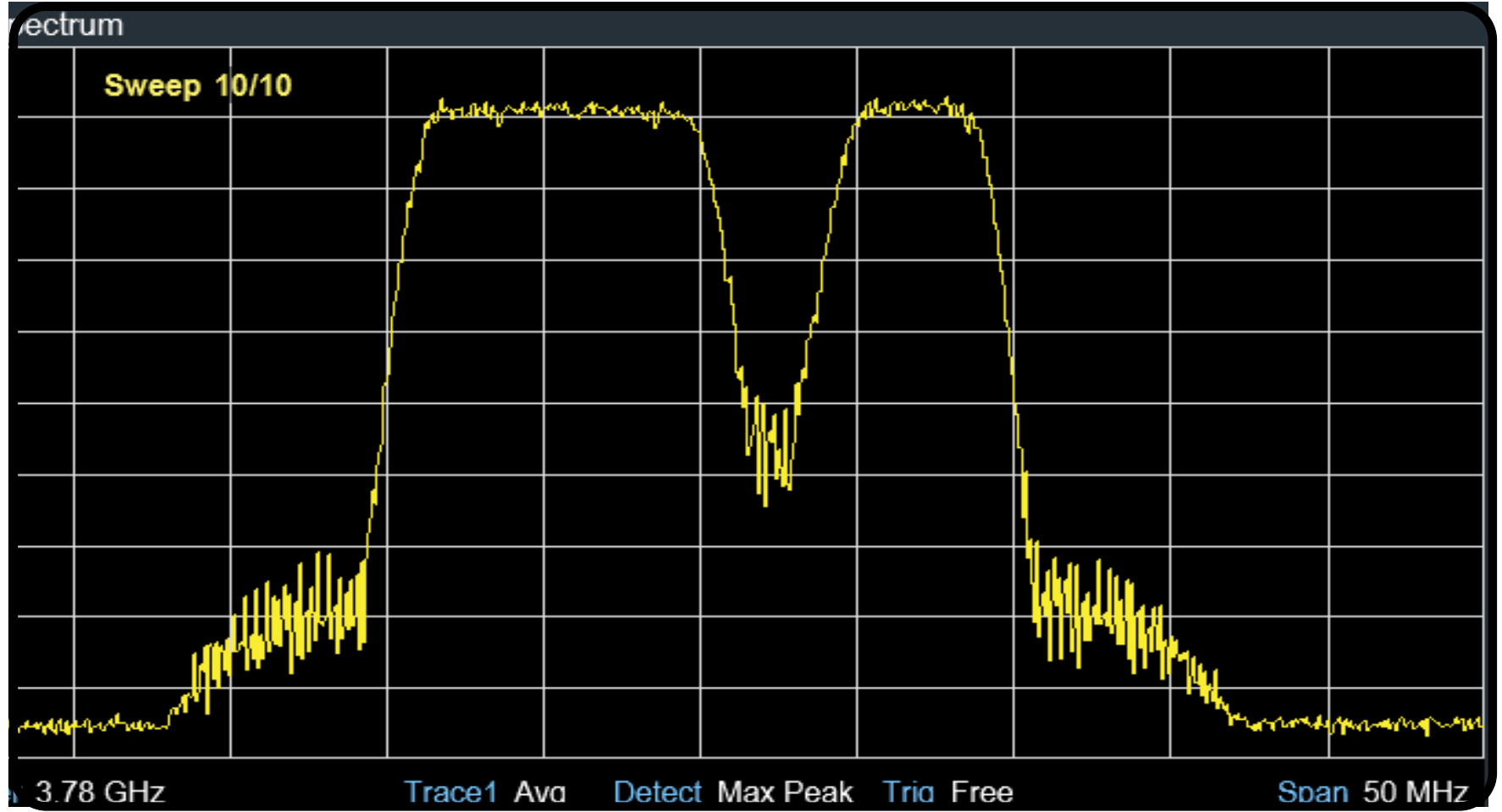}
		\caption[]{\gls{LTE}  at R\&H spectrum analyzer.}
		\label{fig:SpectrumMeasurment2b}
    \end{subfigure}    
    \begin{subfigure}{.31\textwidth}
        \centering
		\includegraphics[width=1\linewidth]{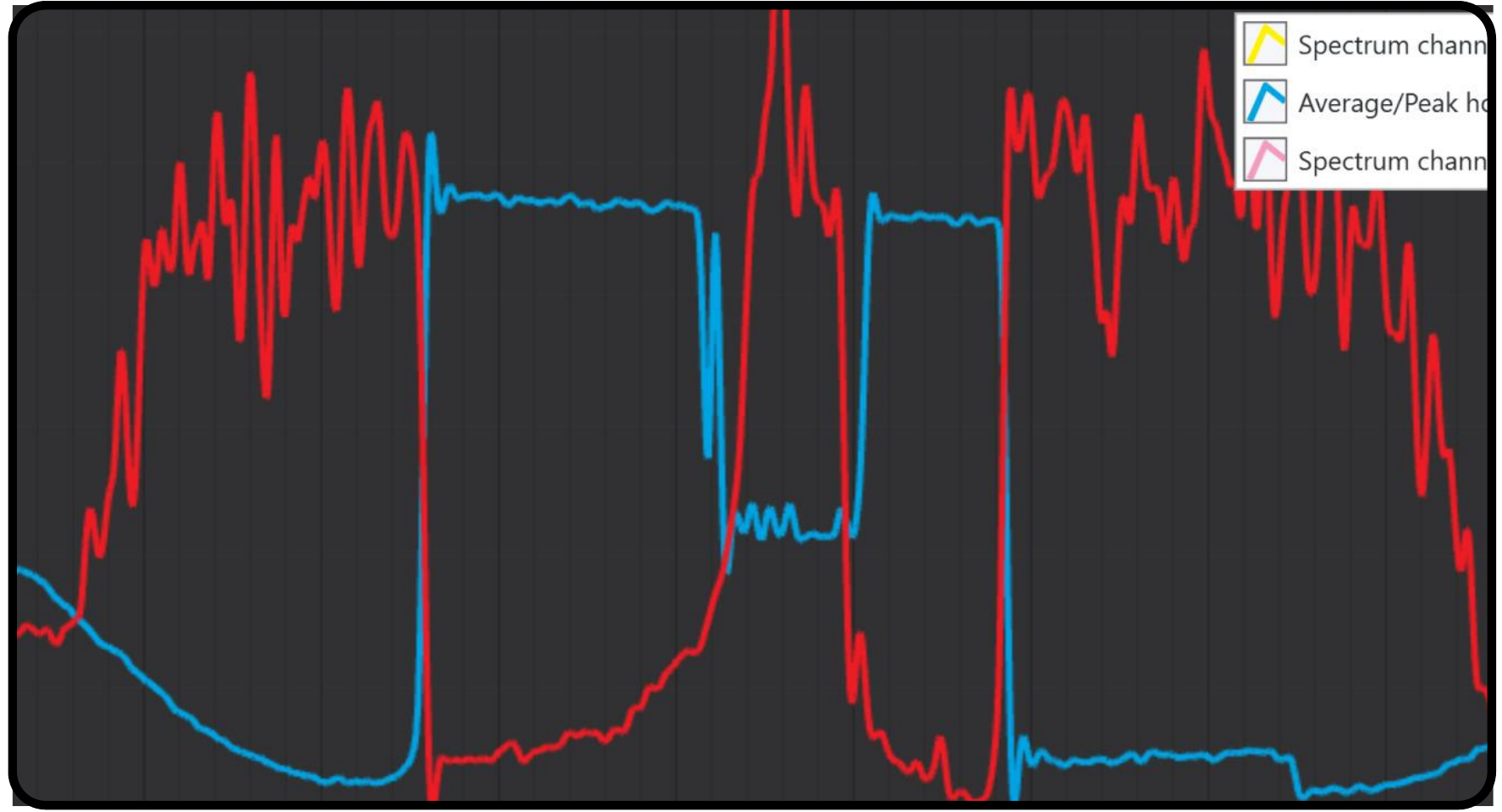}
		\caption[]{Radar and \gls{LTE} at developed application.}
		\label{fig:SpectrumMeasurment2c}
    \end{subfigure}        
    \caption[]{Screen captures of the  resulting spectrum occupied by the \gls{LTE} communications and radar signals  at the developed two-channel spectrum sensing application and R\&H spectrum analyzer. The spectrum of the \gls{LTE} downlink in (a) is validated by a commercial product in (b), and (c) indicates the the  resulting spectrum of both communications (blue) and radar (red) at the developed two-channel spectrum sensing application. 
    }
    \label{fig:SpectrumMeasurment2}
\end{figure*}

According to these tables, we utilize the radar with a $50 \%$ duty cycle. By transmitting a set of $M = 2$ waveforms with length $N = 400$, radar transmissions will occupy a bandwidth of $40$ MHz. On the other side, the \gls{LTE} communications framework utilizes $20$ MHz bandwidth for  transmission. 
Further, we select the allocation $1111111111110000000111111$ for the LTE resource block ($4$ physical resource blocks/bit), where the entry ``1''  indicates the use of the corresponding time-bandwidth resources in the \gls{LTE} application framework. The spectrum of this \gls{LTE} downlink is measured with the developed spectrum sensing application as depicted in \figurename{~\ref{fig:SpectrumMeasurment2a}}. By setting $\theta = 1$ in \eqref{eq:sum_weighted}, the radar optimizes its transmit waveform to avoid interference with the bands occupied by the \gls{LTE} resource blocks, while exploiting the holes in the $20$ MHz of the \gls{LTE} spectrum. { A screen capture of the  resulting spectrum occupied by radar and communications is depicted in \figurename{~\ref{fig:SpectrumMeasurment2}}. This figure serves two purposes, (i) focusing on the \gls{LTE} downlink spectrum, it validates the spectrum analyzer application with a commercial product, and 
(ii) it clearly indicates that the desired objective of spectrum shaping is met. The impact of this matching on performance of radar and communications  are presented next.}

\begin{figure*}
    \centering
    \begin{subfigure}{.49\textwidth}
        \centering
		\includegraphics[width=1\linewidth]{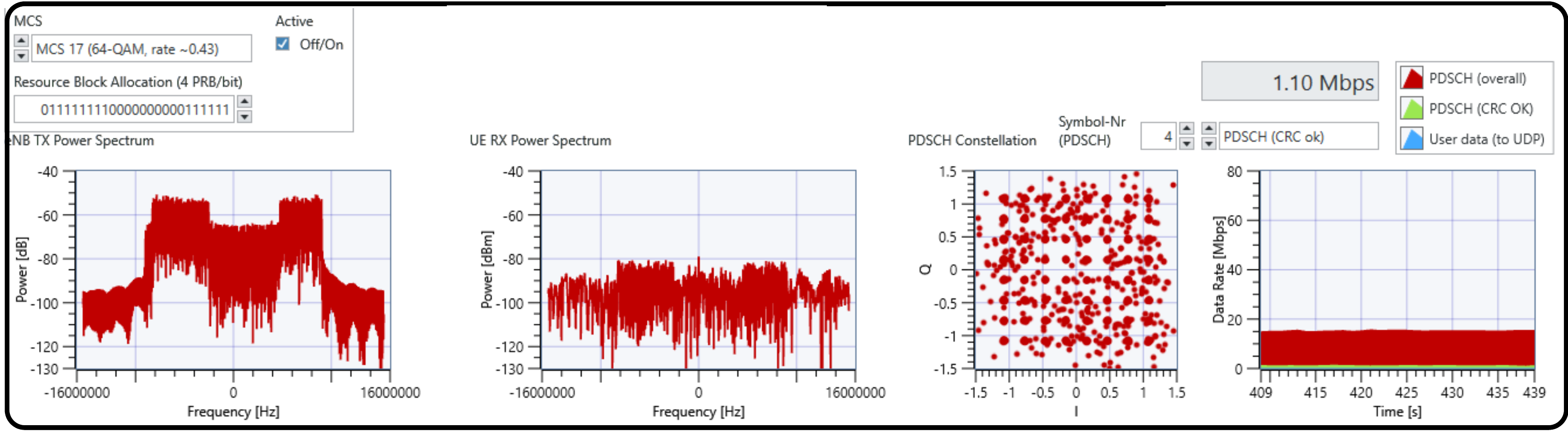}
		\caption[]{Random-phase sequences.}
		\label{fig:LTEandRadara}
    \end{subfigure}
    \begin{subfigure}{.49\textwidth}
        \centering
		\includegraphics[width=1\linewidth]{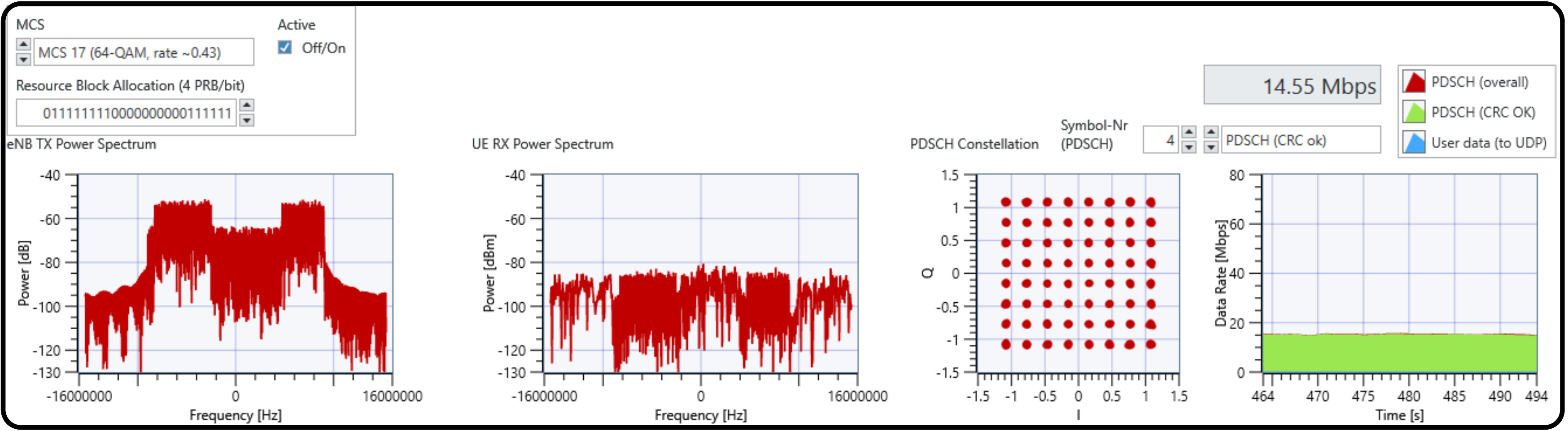}
		\caption[]{Optimized sequences ($\theta = 1$).}
		\label{fig:LTEandRadarb}
    \end{subfigure}   
    \begin{subfigure}{.49\textwidth}
        \centering
		\includegraphics[width=1\linewidth]{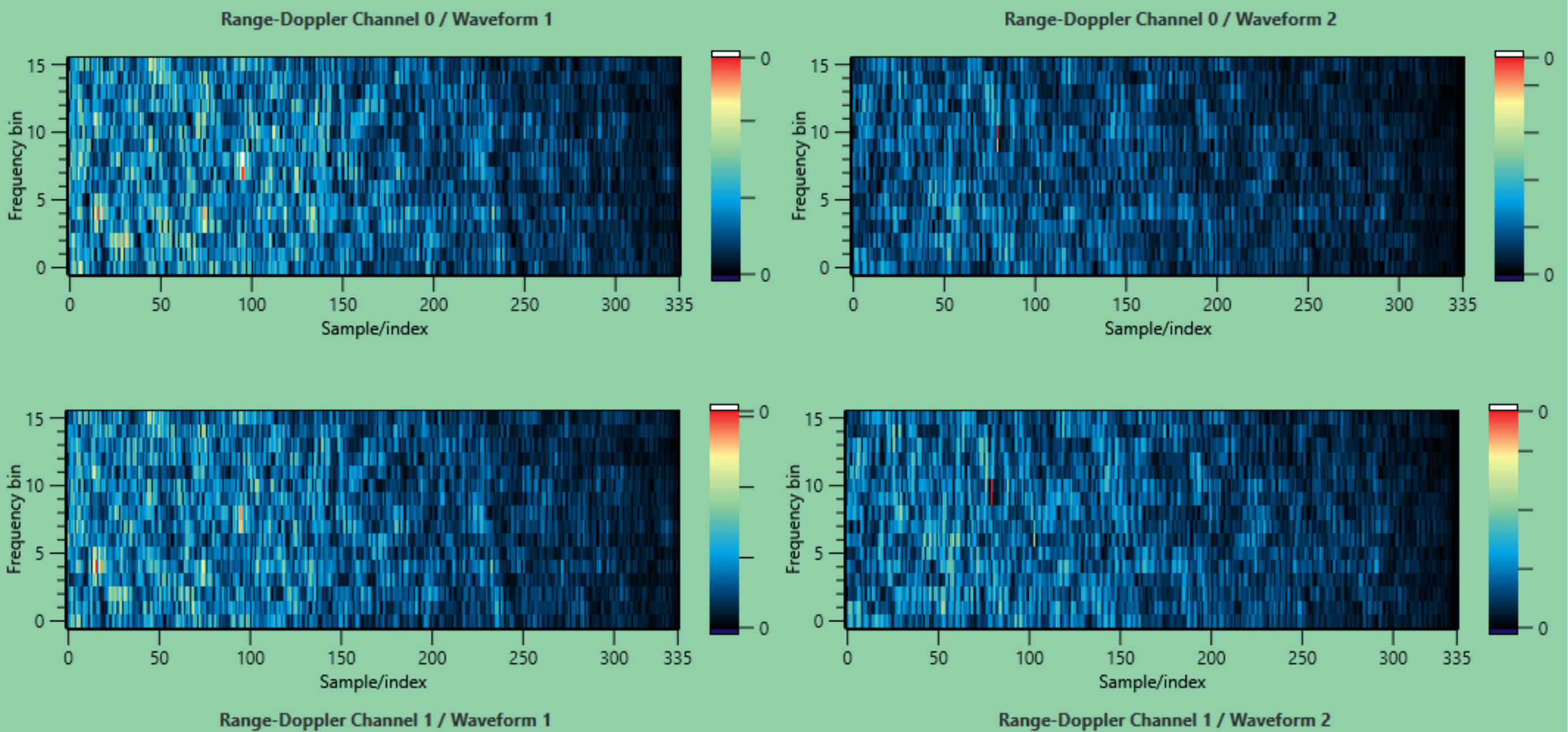}
		\caption[]{Random-phase sequences.}
		\label{fig:LTEandRadarc}
    \end{subfigure}
    \begin{subfigure}{.49\textwidth}
        \centering
		\includegraphics[width=1\linewidth]{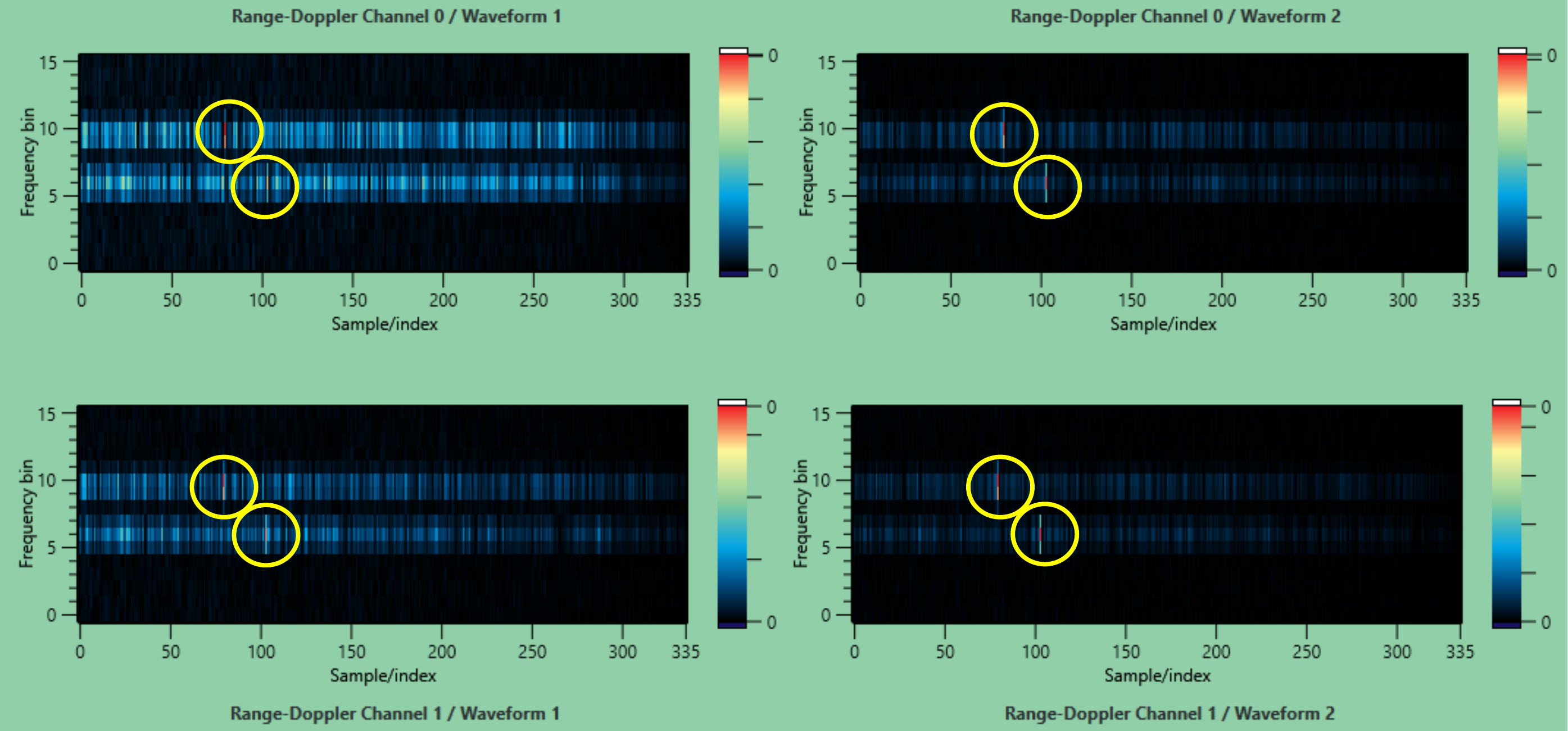}
		\caption[]{Optimized sequences ($\theta = 1$).}
		\label{fig:LTEandRadard}
    \end{subfigure}    
    \caption[]{\gls{LTE} application framework in the presence of radar signal. In case of transmitting random-phase sequences in radar at the same frequency band of communications, the throughput of communications decreases drastically which is depicted in (a). In this case, radar also cannot detect targets as depicted in (c). In case of transmitting the optimized waveforms, the performance of both radar and communications enhances (b and c).}
    \label{fig:LTEandRadar}
\end{figure*}

When the radar transmits random-phase sequences, it utilizes the entire bandwidth and the two system mutually interfere. In fact, the operations of both radar and communications are disrupted as depicted in \figurename{~\ref{fig:LTEandRadar}} (a and c),  thereby creating difficulties for their coexistence. In this case, by utilizing the \gls{SILR} optimized waveforms obtained by setting $\theta = 1$ in \textbf{Algorithm \ref{alg:waveform_design}}, the performance of both systems are enhanced as indicated pictorially  in \figurename{~\ref{fig:LTEandRadar}} (b and d). A possible short-coming is the potentially higher cross-correlation levels of the two transmit waveforms leading to false targets impacting the radar performance.  
\subsection{ICCL Minimization}
\begin{figure}
    \centering
    \begin{subfigure}{.45\textwidth}
        \centering
		\includegraphics[width=1\linewidth]{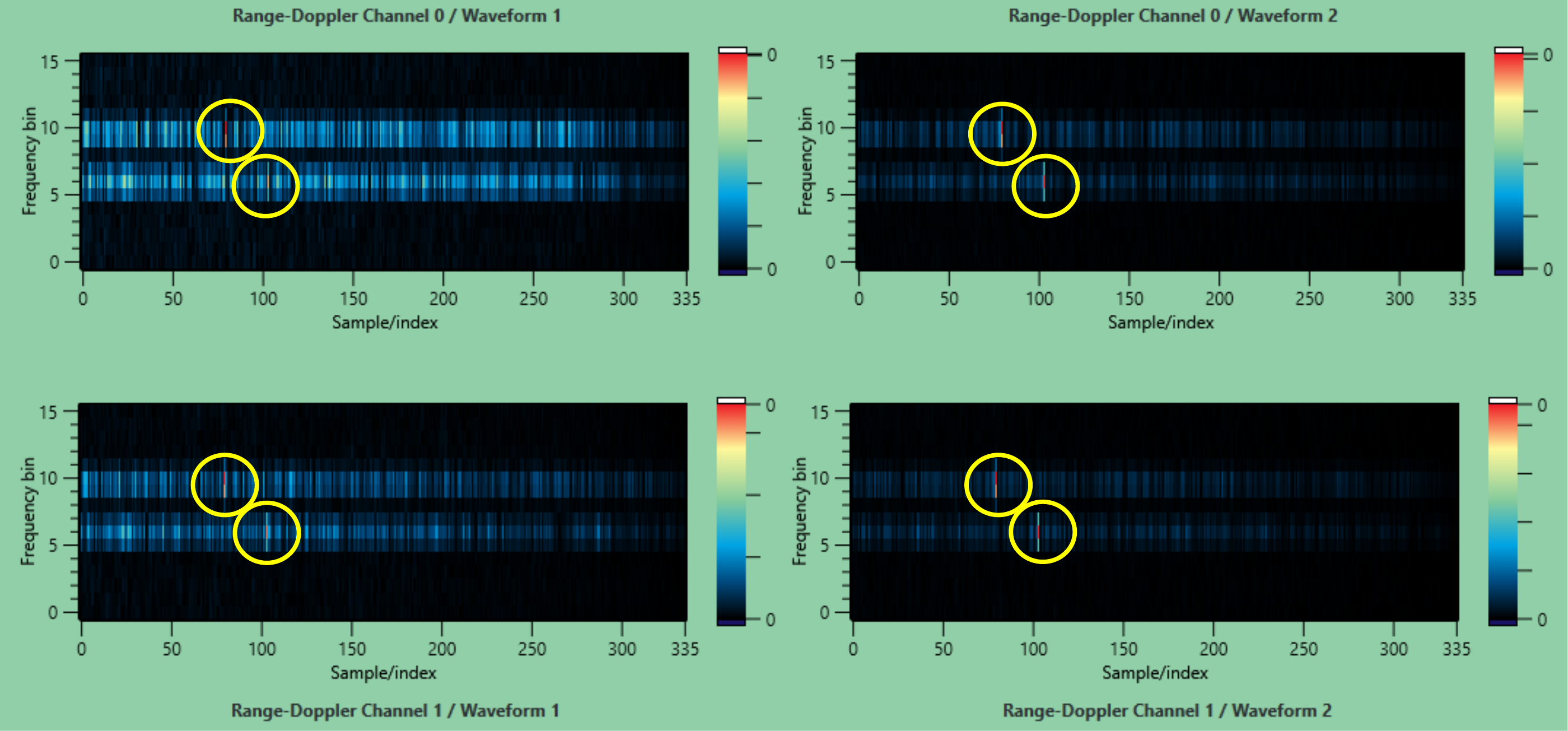}
		\caption[]{Random-phase sequences.}
		\label{fig:radarICCLa}
    \end{subfigure}
    \begin{subfigure}{.45\textwidth}
        \centering
		\includegraphics[width=1\linewidth]{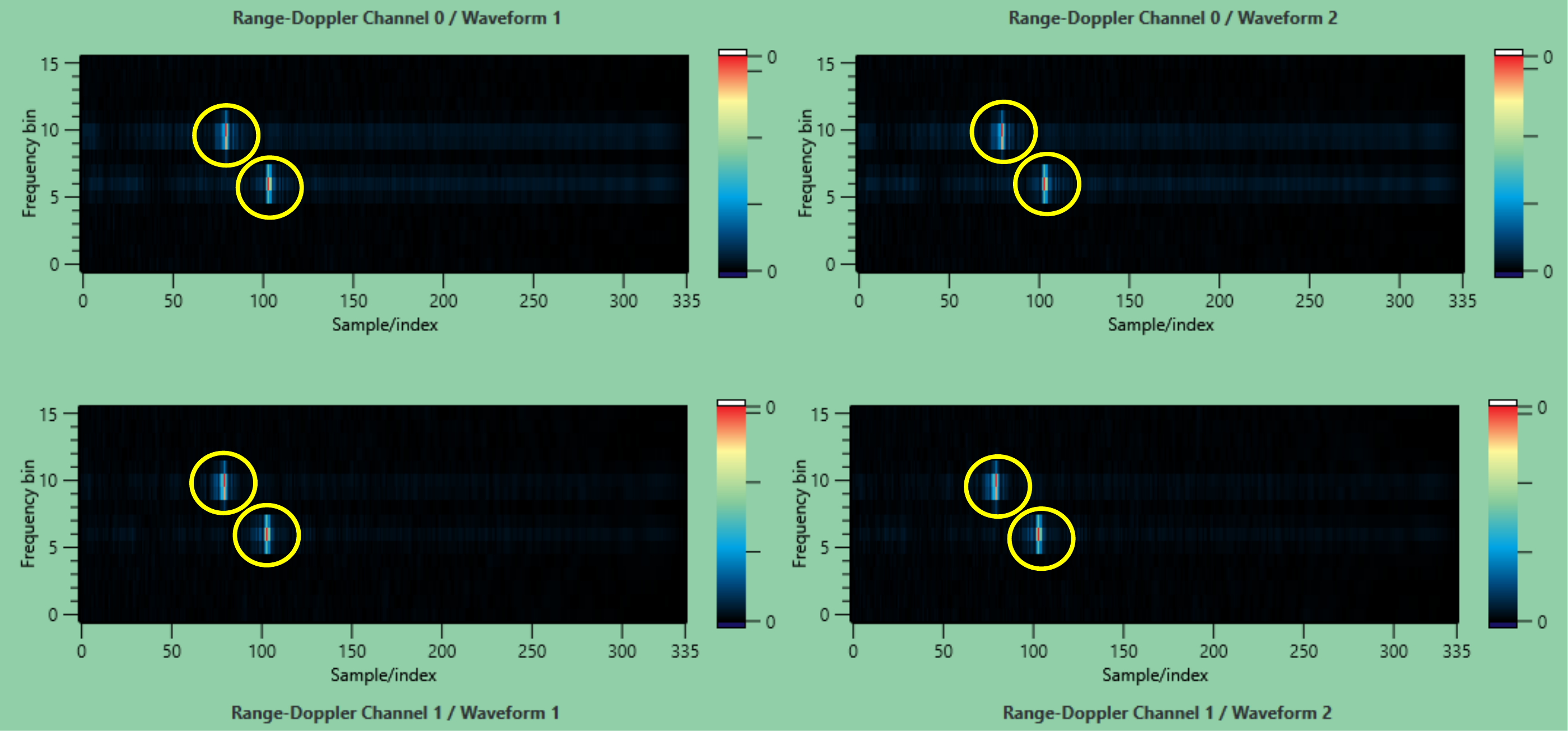}
		\caption[]{Optimized waveforms ($\theta = 0$).}
		\label{fig:radarICCLb}
    \end{subfigure}
    \caption[]{Range-Doppler plots for random-phase and optimized sequences obtained by \textbf{Algorithm \ref{alg:waveform_design}} with $\theta = 0$.}
    \label{fig:radarICCL}
\end{figure}

In the absence of the \gls{LTE} downlink, the radar can optimize its waveform based on \gls{ICCL} minimization by setting $\theta = 0$ in \textbf{Algorithm \ref{alg:waveform_design}}. \figurename{~\ref{fig:radarICCL}} indicates the performance of the radar system in this case in comparison with a set of random-phase sequences. The lower cross-correlation of the optimized sequences can be visually observed from the lack of sidelobes in this figure.  { Since the \gls{SILR} is not considered, this optimized sequence is observed to occupy the entire band causing degradation to communication link}.
%
\subsection{Trade-off between SILR and ICCL}
\begin{figure}
    \centering
	\includegraphics[width=0.92\linewidth]{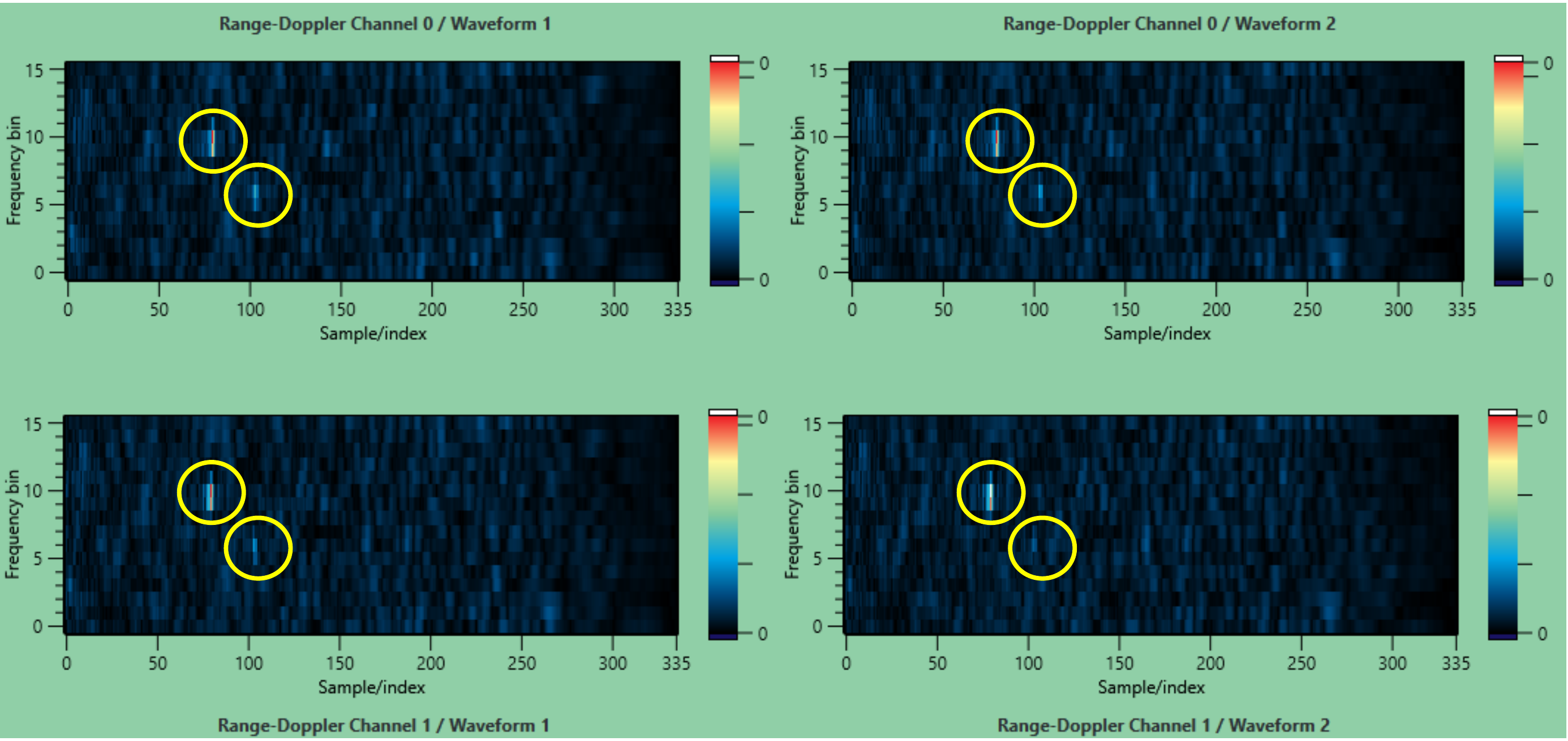}
    \caption[]{The performance of the radar in case of transmitting optimized waveforms with $\theta = 0.75$.}
    \label{fig:radarOptimizedICCLSLIR}
\end{figure}

This section evaluates the performance of the optimized waveforms when
$\theta = 0.75$, a value shown in Section \ref{Sec:Perfomance} to offer an adequate trade-off between spectrum shaping and low cross-correlation. We first evaluate the performance of radar and communication experiencing mutual interference when the radar does not undertake the waveform optimization. Subsequently, we let the radar optimize its waveform by setting $\theta = 0.75$. 
\figurename{~\ref{fig:radarOptimizedICCLSLIR}} depicts range-Doppler plots of this case.  
To evaluate the corresponding performances,  we calculate the \gls{SINR} of the two targets for radar while on the communication side, report the \gls{PDSCH} throughput calculated by the \gls{LTE} application framework. We perform our experiments in following steps: 
\begin{itemize}
	\item Step-1: In the absence of radar transmission, we collect the \gls{LTE} \gls{PDSCH} throughput for MCS0, MCS10 and MCS 17. For each \gls{MCS}, we use \gls{LTE} transmit power of $5$ dBm, $10$ dBm, $15$ dBm and $20$ dBm.   
	\item Step-2: In the absence of \gls{LTE} transmission, we obtain the received \gls{SNR} for the two targets. The \gls{SNR} is calculated as the ratio of the peak power of the detected targets to the average power of the cells close to the target location in the range-Doppler map.
	\item Step-3: We transmit a set of random-phase sequences as the radar waveform. At the same time, we transmit the \gls{LTE} waveform and let the two waveforms interfere with each other. We log the \gls{PDSCH} throughput as well as the \gls{SINR} of Target-$1$ and Target-$2$. We perform this experiment for MCS0, MCS10 and MCS17 and for each \gls{MCS}, we increase the \gls{LTE} transmit power from $5$ dBm to $20$ dBm in steps of $5$ dBm. Throughout the experiment we keep the radar transmit power fixed. For each \gls{LTE} \gls{MCS} and \gls{LTE} transmit power combination, we average over $5$ experiments before logging the \gls{PDSCH} throughput and target \gls{SINR}s.    
	\item Step-4: We repeat step-3, 
	but using the optimized waveforms with $\theta = 0.75$ at the radar transmitter. This optimization is carried out in real-time as mentioned earlier using the spectrum sensing application and the \textbf{Algorithm \ref{alg:waveform_design}} implementation.
\end{itemize} 
\begin{figure*}
	\centering
	\begin{subfigure}{.32\textwidth}
		\centering
		\includegraphics[width=1\linewidth]{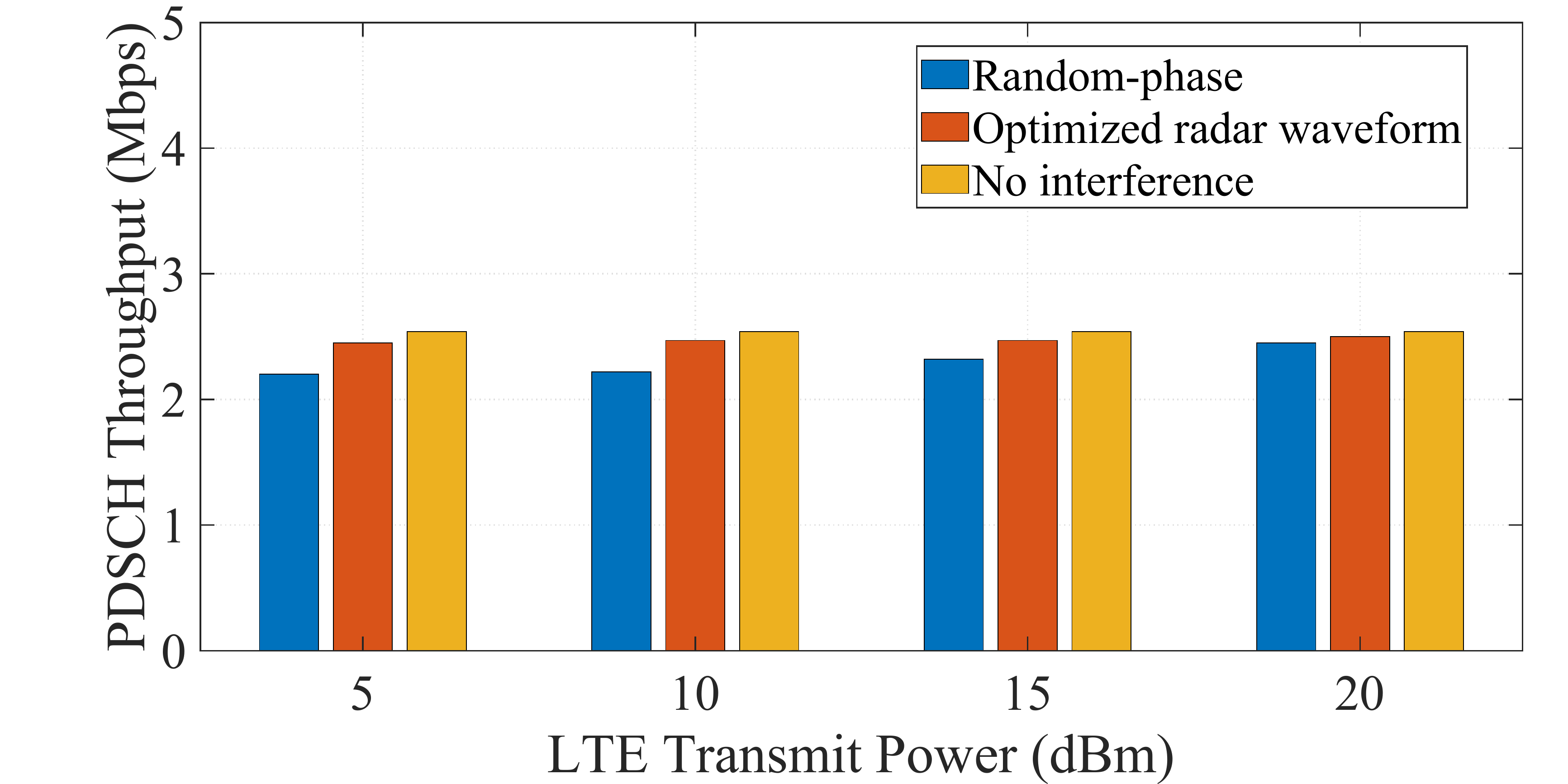}
		\caption[]{MCS 0 (QPSK $0.12$)}\label{fig:lte_mcs0}
	\end{subfigure}
	\begin{subfigure}{.32\textwidth}
		\centering
		\includegraphics[width=1\linewidth]{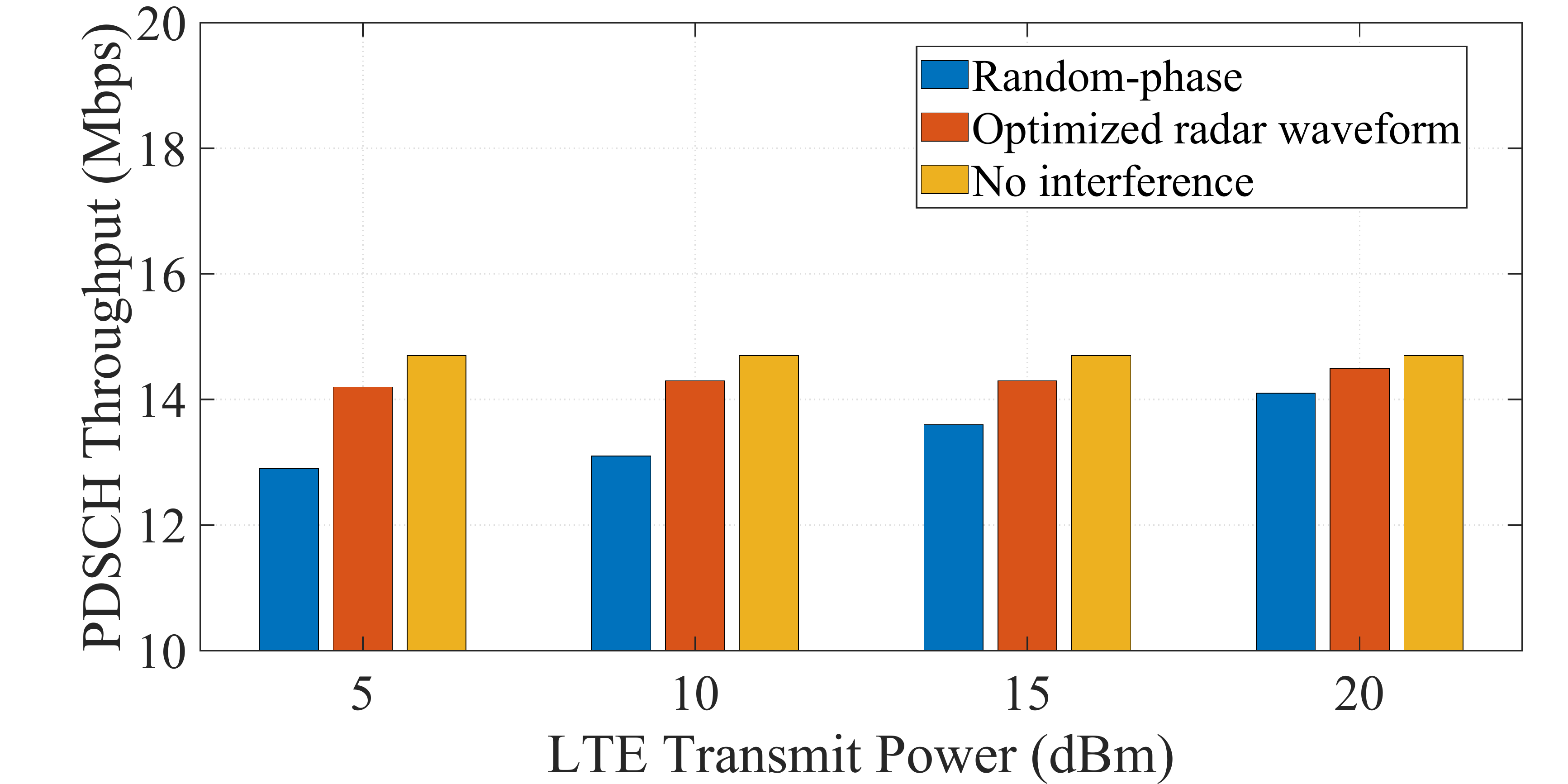}
		\caption[]{MCS 10 (16QAM $0.33$)}\label{fig:lte_mcs10}
	\end{subfigure}
	\begin{subfigure}{.32\textwidth}
		\centering
		\includegraphics[width=1\linewidth]{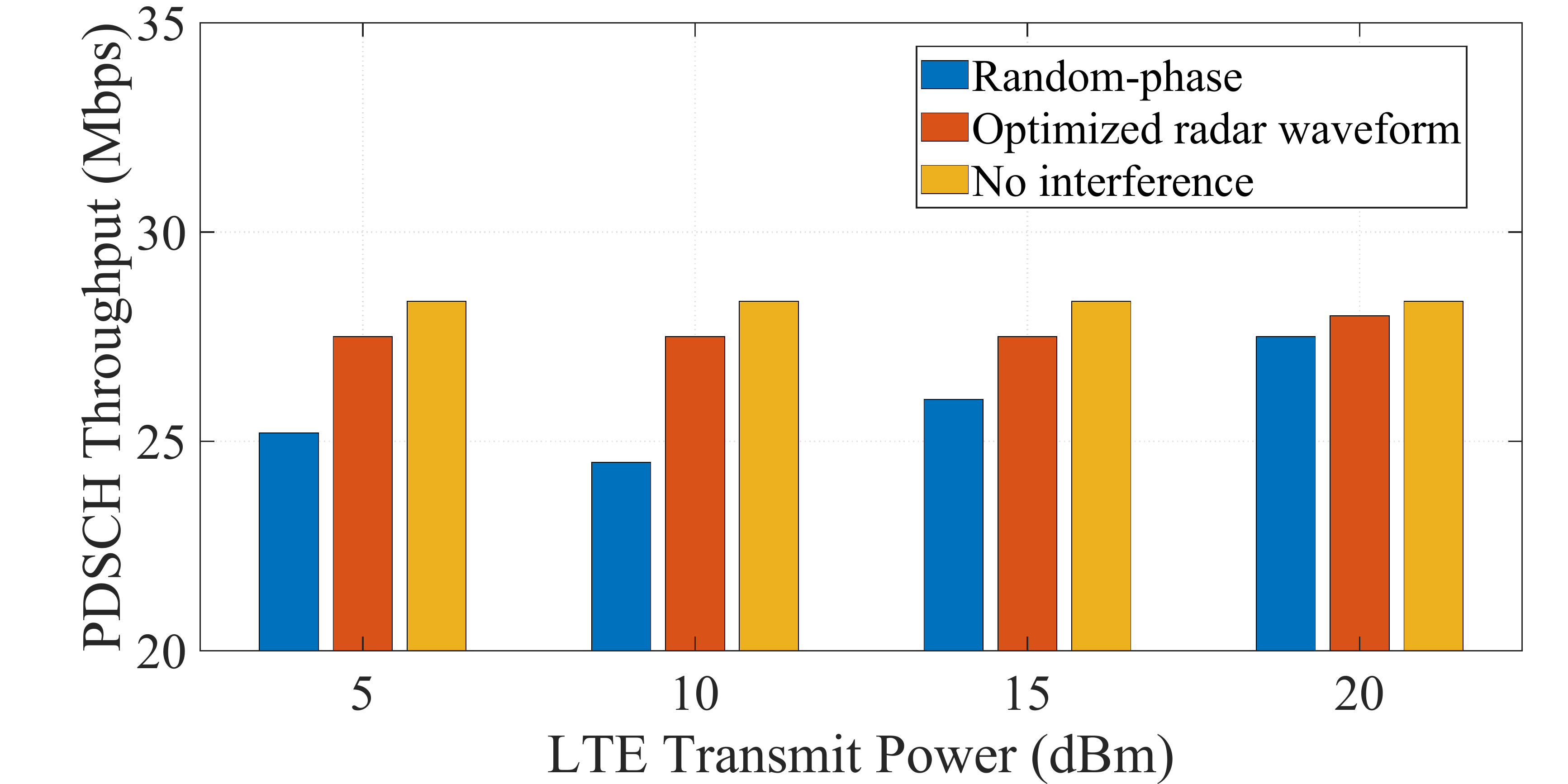}
		\caption[]{MCS 17 (64QAM $0.43$)}\label{fig:lte_mcs17}
	\end{subfigure}
	\caption[]{PDSCH throughput of \gls{LTE} under radar interference. We observe that with radar interference reduces the \gls{PDSCH} throughput but with cognitive spectrum sensing followed by spectral shaping of the radar waveform \gls{PDSCH} throughput improves for all the \gls{LTE} \gls{MCS}.
	}\label{fig:lte-performance}
\end{figure*}


\begin{figure*}
	\centering
	\begin{subfigure}{.32\textwidth}
		\centering
		\includegraphics[width=1\linewidth]{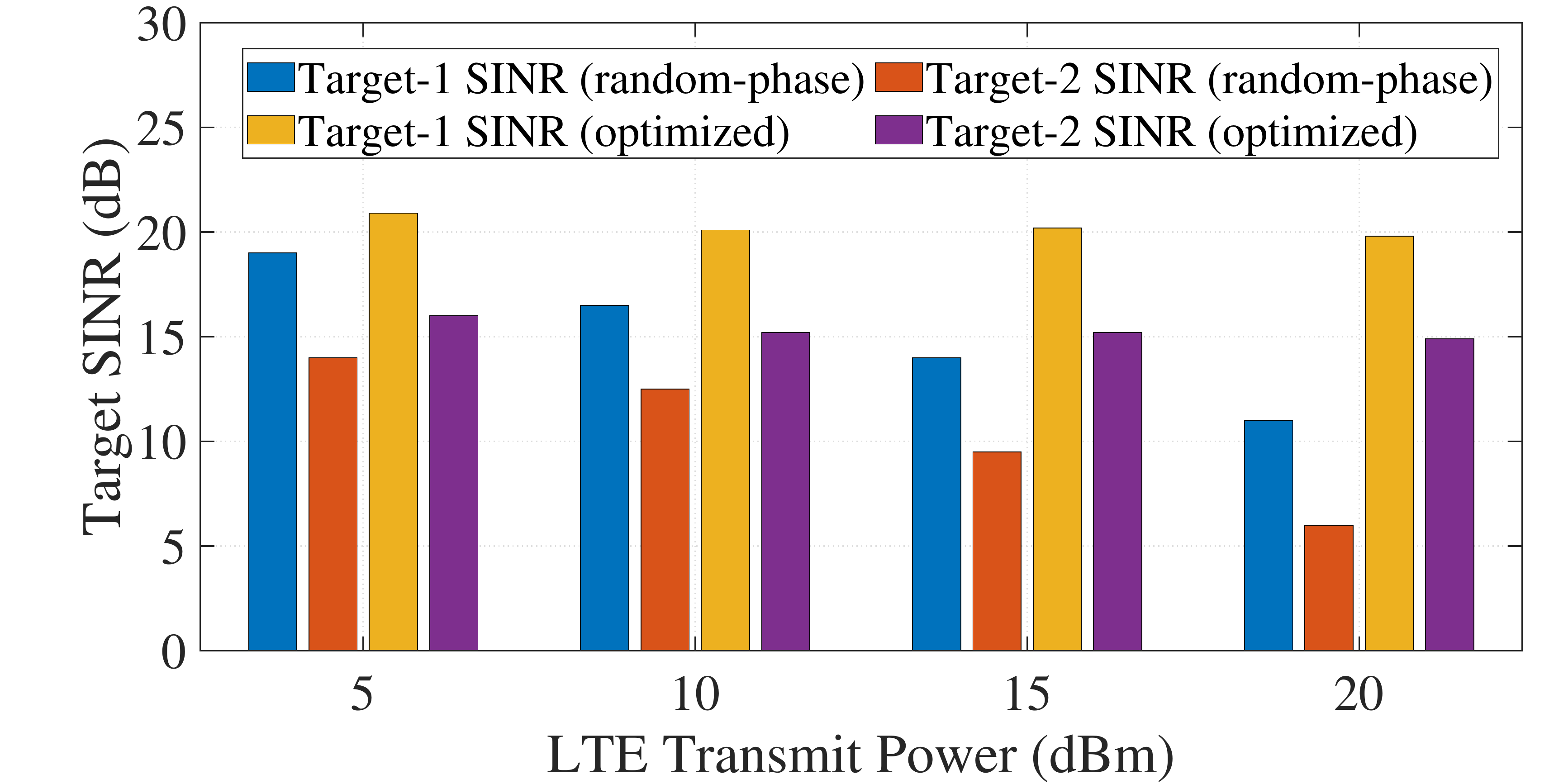}
		\caption[]{MCS 0 (QPSK $0.12$)}\label{fig:radar-performancelte_mcs0}
	\end{subfigure}
	\begin{subfigure}{.32\textwidth}
		\centering
		\includegraphics[width=1\linewidth]{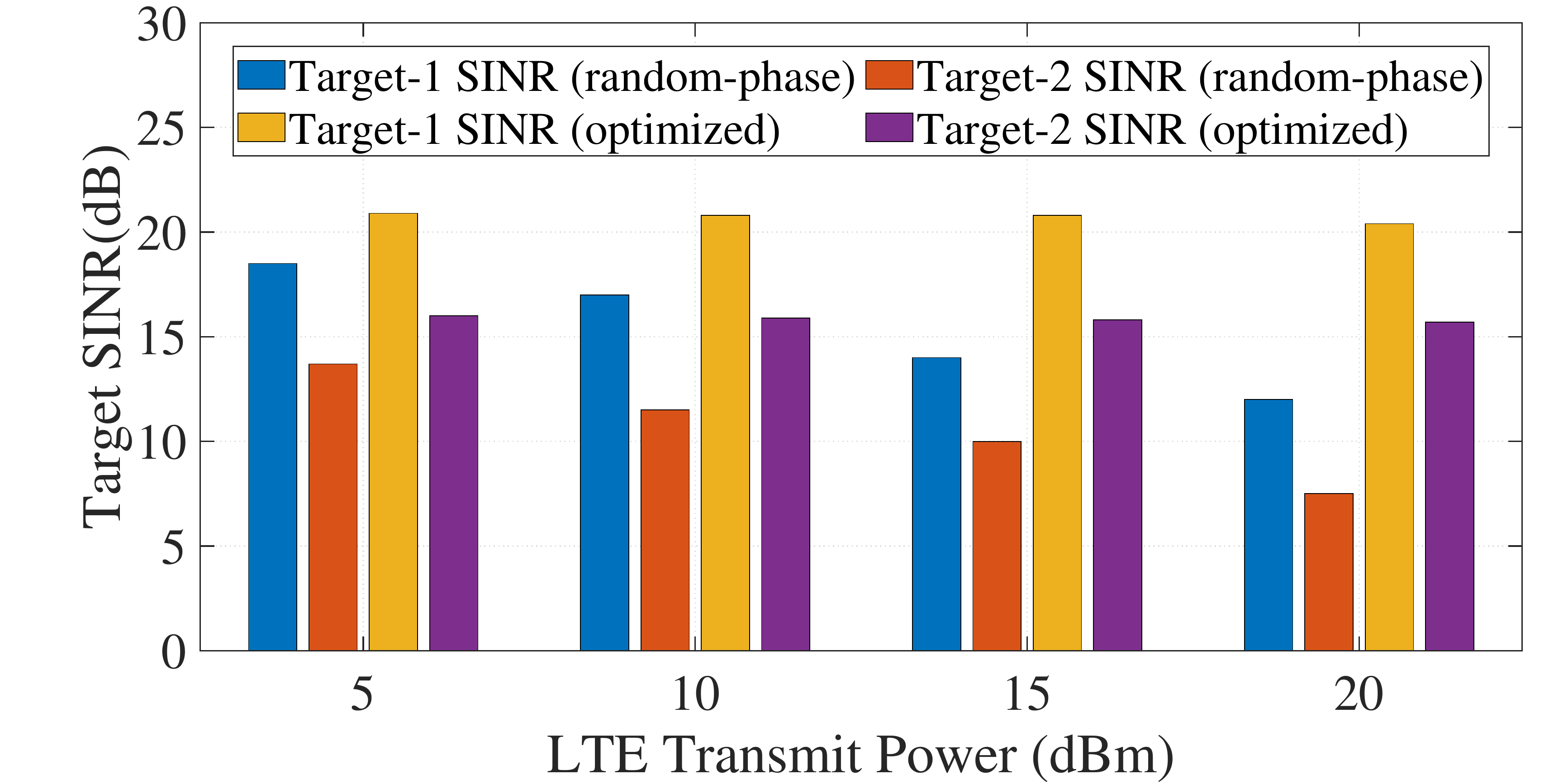}
		\caption[]{MCS 10 (16QAM $0.33$)}\label{fig:radar-performancelte_mcs10}
	\end{subfigure}
	\begin{subfigure}{.32\textwidth}
		\centering
		\includegraphics[width=1\linewidth]{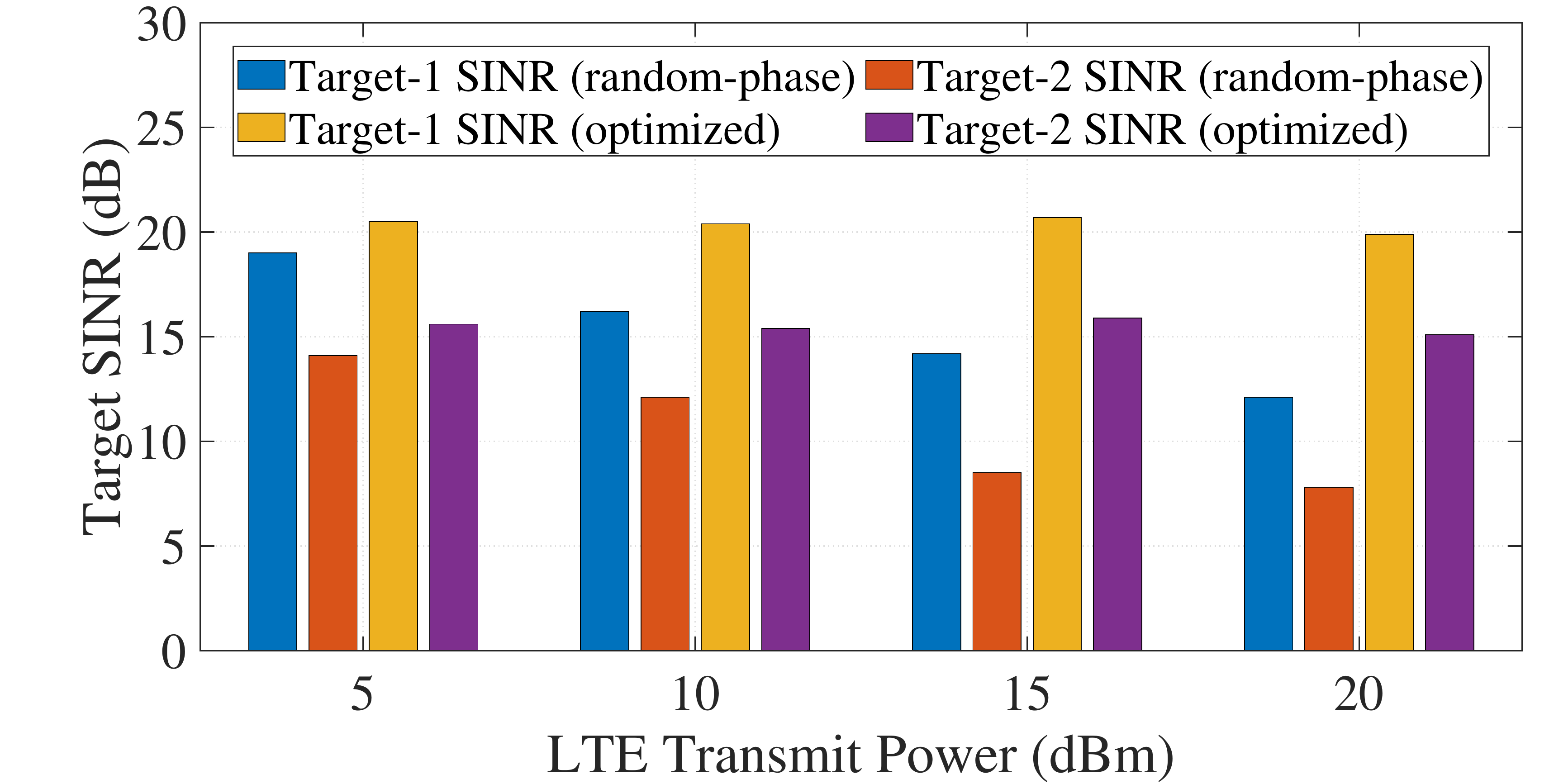}
		\caption[]{MCS 17 (64QAM $0.43$)}\label{fig:radar-performancelte_mcs17}
	\end{subfigure}
	\caption[]{\gls{SINR} of targets under interference from donwlink \gls{LTE} link. We observe that by optimizing the transmitting waveforms, the \gls{SNR} of both the targets improves. Note that in this experiment the \gls{SNR} upper-bound for the first and the second target in the absence of communications interference was $22$ dB, and $17$ dB, respectively.  }\label{fig:radar-performance}
\end{figure*}

The performance of the communication link in terms of \gls{PDSCH} throughput is shown in \figurename{~\ref{fig:lte-performance}}. We first observe that, the throughput of the link degrades in the presence of radar interference. The degradation becomes prominent at higher MCS since the \gls{SINR} requirement to obtain a clean constellation for larger modulations is also high. Subsequently, when the radar optimizes its waveform as per \textbf{Algorithm \ref{alg:waveform_design}}, the \gls{LTE} throughput improves. 
Again we see that the improvement is prominent in the higher MCS.   
This is due to the fact that after a certain \gls{SINR}, the lower MCS do not show any symbol error as the constellation points are already well separated. But as the distance between the constellation points decrease, even a small improvement of \gls{SINR} leads to improved Error Vector Magnitude (EVM) which leads to improved decoding and hence a prominent increase in throughput.       

We set $\theta = 0.75$, such that the optimized waveforms avoid the frequency bands occupied by communications, and simultaneously have an acceptable cross-correlation. In this case, the performance of the radar in terms of target \gls{SNR} is shown in \figurename{~\ref{fig:radar-performance}}. In the presence of \gls{LTE} interference, we observe that the \gls{SINR}s of Target-1 and Target-2 degrade.  These quantities improve when the radar optimizes the transmitting waveforms.
Interestingly, when the \gls{LTE} transmission power is high ($15$ dBm, and $20$ dBm), higher  improvement results from the avoidance of the used \gls{LTE} bands. Precisely, when the communication system is transmitting with a power of $20$ dBm, use of the optimized waveforms enhances the \gls{SINR} of Target-1, and Target-2 in excess of $7$ dB in all the MCS values. Note that, due to the different attenuation paths that is considered for the two targets (see \tablename{~\ref{tab:radar_param}} ), the measured \gls{SINR}s for these targets are different. Also, in the absence of the \gls{LTE} interference, the achieved \gls{SINR} of Target-1, and Target-2 is $22$ dB and $17$ dB, respectively, {which is the upper bound for the achievable \gls{SINR} through the  optimized waveforms in presence of the communications interference.} 

\section{Conclusion}\label{Sec:Conclusion}
In this paper, a radar-centric approach has been pursued towards a coexistence with communication systems. Particularly, the paper developed a  \gls{SDR} based cognitive \gls{MIMO} radar prototype using \gls{USRP} devices that coexist with \gls{LTE} links. To enable seamless operation of incumbent \gls{LTE} links and opportunistic radar sensing, the paper relied on cognition achieved through the implementation of a spectrum sensing followed by the development of a \gls{MIMO} waveform design process. An  algorithm based on \gls{CD} approach is considered to design  a set of sequences, where the  optimization is based on real-time feedback received from the environment through the spectrum sensing application. The developed prototype is tested both in controlled environment and \gls{OTA} to validate its functionalities and the experimental results indicating adherence to system requirements and performance enhancement are noted. A further development of this first coexistence prototype can be spatial beamforming in addition to the spectral shaping that can be considered in advanced cognitive \gls{MIMO} radar systems. 
%
\appendices
\section{}\label{app:1}
\paragraph{\gls{SILR} coefficients} Let $g_a(\bX)$ and $g_b(\bX)$ be the numerator and denominator of $g_s(\bX)$ respectively. Therefore $g_a(\bX)$ can be written as,
\par\noindent\small
\begin{equation}
\begin{aligned}
    g_a(\bX) \triangleq& \textstyle \sum_{m=1}^{M} \norm{\mathbf{f}_k^{\dagger} \mathbf{x}_m}^2 | k \in \mathcal{U} = \sum_{m=1}^{M} \sum_{k \in \mathcal{U}} \mathbf{x}_m^{\dagger} \mathbf{f}_k \mathbf{f}_k^{\dagger} \mathbf{x}_m \\
    =&\textstyle \sum_{m=1}^{M} \mathbf{x}_m^{\dagger} \bF_{\mathcal{U}} \mathbf{x}_m = \sum_{m=1}^{M}\sum_{n=1}^{N}\sum_{l=1}^{N} x_{m,n}^*u_{n,l}x_{m,l} 
\end{aligned}
\end{equation}
\normalsize
where, $\bF_{\mathcal{U}} \triangleq \sum_{k \in \mathcal{U}} \mathbf{f}_k \mathbf{f}_k^{\dagger} \in \mathbb{C}^{N \times N}$ and $u_{n,l}$ are the elements of matrix $\bF_{\mathcal{U}}$. By some mathematical manipulation it can be shown that, $g_a(\bX) = a_0x_{t,d} + a_1 + a_2x_{t,d}^*$, where, 
\par\noindent\small
\begin{equation} 
\begin{aligned}
    a_0 &\triangleq \textstyle \sum_{\substack{{n = 1}\\{n \neq d}}}^{N}x^*_{t,n}u_{n,d}, \ a_2 \triangleq a_0^*, \\
    a_1 &\triangleq \textstyle \sum_{\substack{{n = 1}\\{n \neq d}}}^{N}\sum_{{n,l = 1}}^{N} x^*_{m,n}u_{n,l}x_{m,l} + \sum_{\substack{{n,l = 1}\\{n,l \neq d}}}^{N} x^*_{t,n}u_{n,l}x_{t,l} + u_{d,d}.
\end{aligned}
\end{equation}
\normalsize
Let us assume that, $\bF_{\mathcal{V}} \triangleq \sum_{k \in \mathcal{V}} \mathbf{f}_k \mathbf{f}_k^{\dagger} \in \mathbb{C}^{N \times N}$. Likewise, $g_b(\bX) = b_0x_{t,d} + b_1 + b_2x_{t,d}^*$, where $b_i$'s are obtained similar to $a_i$ with $v_{n,d}$, $v_{n,l}$ and $v_{d,d}$ replacing $u_{n,d}$, $u_{n,l}$ and $u_{d,d}$ respectively, and $v_{n,l}$ are the elements of matrix $\bF_{\mathcal{V}}$.

\paragraph{\gls{ICCL} coefficients} The \gls{ICCL} can be written as, $\tilde{g}_c(\bX) = \sum_{\substack{{m=1}\\{m \neq t}}}^{M}\sum_{\substack{{m'=1}\\{m' \neq m,t}}}^{M}\sum_{k=-N+1}^{N-1}|r_{m,m'}(l)|^2 + \sum_{\substack{{m'=1}\\{m' \neq t}}}^{M}\sum_{l=-N+1}^{N-1}|r_{t,m'}(l)|^2 + \sum_{\substack{{m=1}\\{m \neq t}}}^{M}\sum_{l=-N+1}^{N-1}|r_{m,t}(l)|^2$. Since $\norm{r_{t,m}(l)}_2^2 = \norm{r_{m,t}(l)}_2^2$, hence, $\tilde{g}_c(\bX)$ becomes,
\begin{equation}
	\tilde{g}_c(\bX) = \gamma_t + 2\textstyle \sum_{\substack{{m=1}\\{m \neq t}}}^{M}\sum_{l=-N+1}^{N-1}|r_{m,t}(l)|^2
\end{equation}
where, $\gamma_t \triangleq \sum_{\substack{{m=1}\\{m \neq t}}}^{M}\sum_{\substack{{m'=1}\\{m' \neq m,t}}}^{M}\sum_{k=-N+1}^{N-1}|r_{m,m'}(l)|^2$. Then $r_{m,t}(l)$ can be written as, $r_{m,t}(l) = \alpha_{mtdl}x_{t,d} + \gamma_{mtdl}$, where \cite{8706639},
\par\noindent\small
\begin{equation}
	\gamma_{mtdl} \triangleq \textstyle \sum_{\substack{{n=1}\\{n \neq d-k}}}^{N-k}x_{m,n}s_{t,n+l}^*, \alpha_{mtdl} \triangleq x_{m,d-l}I_A(d-l)
\end{equation}
\normalsize
where, $I_A(p)$ is the indicator function of set $A = \left \{1, \dots, N\right \}$. By some mathematical manipulation the $g_c(\bX)$ can be expressed as, $g_c(\bX) = c_0 x_{t,d} + c_1 + c_2 x_{t,d}^*$, where,
\par\noindent\small
\begin{equation}
\begin{aligned}
c_0 &\triangleq \textstyle\frac{2}{(2MN)^2}\sum_{\substack{{m=1}\\{m \neq t}}}^{M}\sum_{l=-N+1}^{N-1} \alpha_{mtdl}\gamma_{mtdl}^*, \quad c_2 \triangleq c_0^* \\
c_1 &\triangleq \textstyle\frac{1}{(2MN)^2}(\gamma_t + 2\sum_{\substack{{m=1}\\{m \neq t}}}^{M}\sum_{l=-N+1}^{N-1} |\alpha_{mtdl}|^2 \\
&+ \textstyle 2\sum_{\substack{{m=1}\\{m \neq t}}}^{M}\sum_{l=-N+1}^{N-1} |\gamma_{mtdl}|^2).
\end{aligned}
\end{equation}
\normalsize


\ifCLASSOPTIONcaptionsoff
  \newpage
\fi

\bibliographystyle{IEEEtran}
\bibliography{JSTSP_SI_joint_communication_2021}

\end{document}